\documentclass[aps,prb,twocolumn,groupedaddress,showpacs,showkeys]{revtex4}
\usepackage{graphics,graphicx}
\usepackage{color}
\usepackage{amssymb,amsmath}
\usepackage{bookmath}
\usepackage{bm}

\begin{document}
\title{Unconventional superconductors under rotating magnetic field II: \\
thermal transport}
\author{A. Vorontsov}
\author{I. Vekhter}
\affiliation{Department of Physics and Astronomy,
     Louisiana State University, Baton Rouge, Louisiana, 70803, USA}
\date{\today}
\pacs{74.25.Fy, 74.20.Rp, 74.25.Bt} \keywords{anisotropic
superconductors, thermal conductivity, magnetic field}

\begin{abstract}
We present a microscopic approach to the calculations of thermal
conductivity in unconventional superconductors for a wide range of
temperatures and magnetic fields.  Our work employs the
non-equilibrium Keldysh formulation of the quasiclassical theory.
We solve the transport equations using a variation of the
Brandt-Pesch-Tewordt (BPT) method, that accounts for the
quasiparticle scattering on vortices. We focus on the dependence
of the thermal conductivity on the direction of the field with the
respect to the nodes of the order parameter, and discuss it in the
context of experiments aiming to determine the shape of the gap
from such anisotropy measurements. We consider quasi-two
dimensional Fermi surfaces with vertical line nodes and use our
analysis to establish the location of gap nodes in heavy fermion
CeCoIn$_5$ and organic superconductor
$\kappa$-(BEDT-TTF)$_2$Cu(NCS)$_2$.
\end{abstract}
\maketitle

\section{Introduction}

In the preceding paper \cite{AVorontsov:I}, hereafter referred to
as I, we developed a theoretical approach to the vortex state in
unconventional superconductors that allowed us to obtain a closed
form solution for the equilibrium Green's function, and therefore
efficiently compute the density of states and the specific heat
for an arbitrary orientation of the magnetic field. In this work
we extend our approach to the calculation of transport properties,
develop the formalism for computing the electronic thermal
conductivity, and compare our results with experiment.

The rationale for both calculations is to provide theoretical
guidance and  support to continued attempts to establish the
measurements of the anisotropy of the specific heat and thermal
conductivity under rotated field as a leading tool in determining
the structure of the energy gap in unconventional superconductors.
While a number of techniques test the symmetry of the gap via the
surface measurements, the corresponding bulk probes are few.
Semiclassical treatment of the quasiparticle energy in the vortex
state incorporated the Doppler shift due to local value of
superfluid velocity associated with the circulating supercurrents.
This approach predicted that, at low fields, the density of
field-induced states at the Fermi surface oscillates as a function
of the field direction and has a minimum when the applied field is
aligned with the nodal direction, $|\Delta(\vp)|=0$, hence the
suggestion to use the measurements of the low temperature specific
heat to determine the position of nodes \cite{IVekhter:1999R}. The
experiments are quite challenging, and, for now, have been carried
out in few materials
\cite{TPark:2003,TPark:2004,HAoki:2004,TParkReview:2004}.

Variations in the density of states also influence transport
properties, and the measurements of the electronic thermal
conductivity under a rotated field have been used more extensively
to study unconventional superconductors and infer the gap
structure
\cite{FYu:1995,HAubin:1997,TWatanabe:UPd2Al3,KIzawa:CeCoIn5,KIzawa:PrOsSb,KIzawa:BEDT,KIzawa:YNiBCkappa,YMatsuda:2006}.
Experimentally, for a fixed direction of the heat current and
rotated field, the dominant twofold anisotropy is that between the
transport normal to and parallel to the vortices; a much smaller
signal is attributed to the existence of the nodes, see
Ref.~\onlinecite{YMatsuda:2006} for a recent review. Theoretical
analysis of the thermal conductivity is also much more
challenging. There are conceptual difficulties with extending the
``local'' semiclassical approach to calculations of the response
functions, especially for clean systems where the mean free path
exceeds the typical length scale for the variations of the
superfluid velocity, the intervortex distance. Even in moderately
dirty systems, where the use of semiclassical method is justified,
it yields, at best, a local value of the thermal conductivity,
which varies from point to point; consequently the averaging
procedure to obtain the experimentally measured value is far from
obvious. Naive averaging completely misses the twofold anisotropy
\cite{HWon:2005}, and therefore is not trustworthy. The
semiclassical approach does not naturally include the scattering
on the vortices, and attempts to introduce it phenomenologically
\cite{CKubert:1998,IVekhter:2000anis} are promising, but have not
yet lead to a consistent description. Moreover, the experiments on
all but high-T$_c$ and some organic superconductors are done at
fields that are a significant fraction of the upper critical
field, $H_{c2}$, where the accuracy of the semiclassical
approximation may be called into question. Consequently, we argued
that a more microscopic approach is needed.

We use a quasiclassical version of the Brandt-Pesch-Tewordt
approximation \cite{BPT:1967,WPesch:1975}, where the normal
electron part of the matrix Green's function is replaced by its
spatial average over a unit cell of the vortex lattice.
Remarkably, this approximation allows for the closed form solution
for the Green's function that we found in
Refs.~\onlinecite{AVorontsov:2006,AVorontsov:I}, and used, with a
fully self-consistent treatment of the  order parameter and
impurity scattering, to determine the behavior of the specific
heat across the $T$-$H$ phase diagram. Below we review these
results and develop a linear response theory for thermal
transport. Implementation of the approximation in the framework of
transport-like quasiclasical (Eilenberger) equations
\cite{eil68,lar68} ensures that we account for the difference
between single particle and transport lifetimes in scattering off
of vortices: the characteristic intervortex distance is large
compared to lattice spacing, and the scattering on the vortices
corresponds to small momentum transfer, and hence forward
scattering is important. This allows us to treat the twofold
anisotropy (transport normal and parallel to the vortices) on
equal footing with the effect of the nodes, and to develop a
consistent picture of the behavior of the thermal conductivity and
the specific heat under the same assumptions.

As in part I, we consider a quasi-two dimensional Fermi surface to
focus on the comparison with the data on heavy fermion CeCoIn$_5$.
In that materials the specific heat data were interpreted as
supporting the $d_{xy}$ gap symmetry~\cite{HAoki:2004}; however,
as pointed out in our short communication \cite{AVorontsov:2006}
and preceding paper, the anisotropic part of the specific heat
changes sign at moderate fields and temperatures, with maxima,
rather than minima, for the field aligned with the nodes. At low
$T$ and $H$, in the region of validity of the semiclassical
method, our results agree with that of the calculations utilizing
the Doppler shift approach, with minima for the field along the
nodes. Consequently, in light of these observations, we
reinterpreted the results of Ref.~\onlinecite{HAoki:2004} as
possibly supporting the $d_{x^2-y^2}$ gap symmetry. Similar gap
structure was inferred by Izawa et al. using phenomenological
interpretation of the thermal conductivity measurements
\cite{KIzawa:CeCoIn5}. We provide the detailed analysis of the
thermal conductivity here.

\begin{figure}[t]
\centerline{\includegraphics[height=5cm]{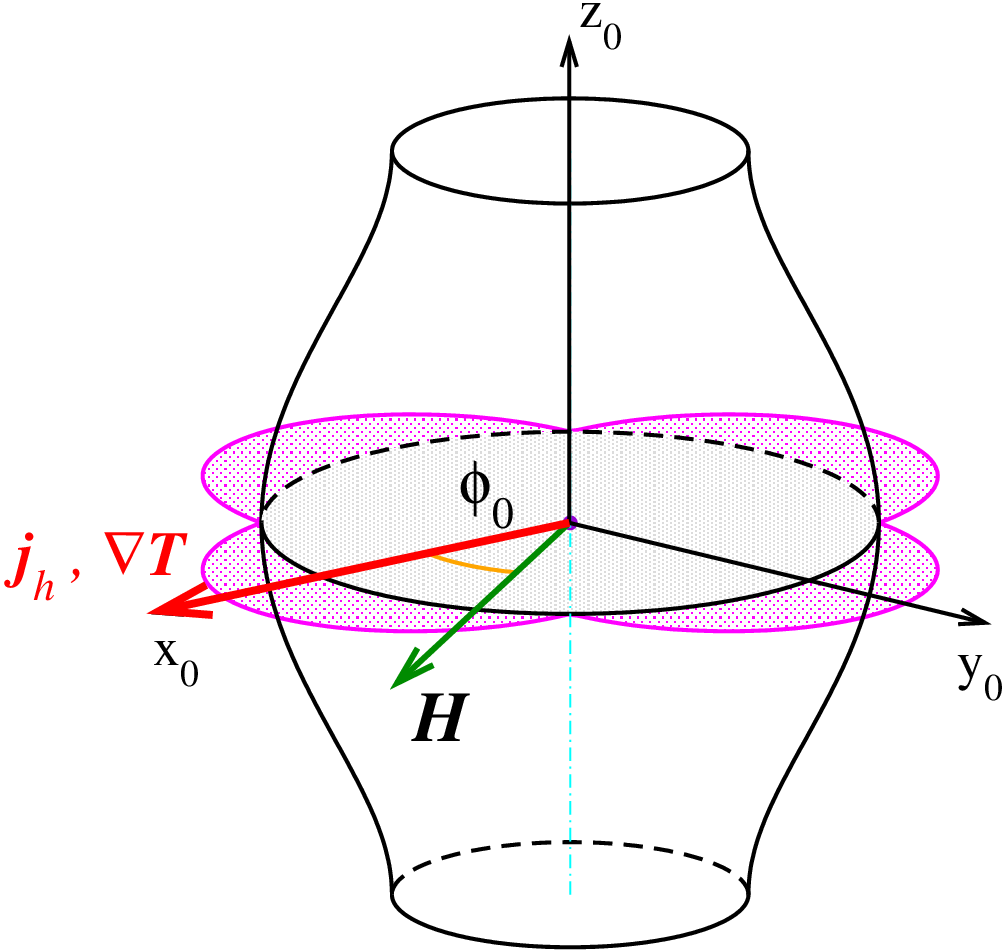}}
\caption{\label{fig:model} (Color online)
The model considered in this paper. Calculation of the
thermal conductivity is done for a quasi-cylindrical Fermi surface,
when the heat current or temperature gradient
and (rotated) magnetic field are in the $ab$-plane.
The order parameter is assumed to have a $d$-wave symmetry.
}
\end{figure}
In the following section we briefly review the approach and the
main results of the preceding paper, such as the closed form
expressions for the Green's functions necessary for computing
linear response to the gradient of temperature.
Sec.~\ref{sec:Thermo} gives the derivation of the thermal
conductivity using Keldysh formulation for the non-equilibrium
theory of superconductivity, with some details relegated to
Appendix~\ref{app:Keldysh}. As in I, we find that the simple
example of a 2D $d$-wave superconductor with a cylindrical Fermi
surface provides a semi-analytically accessible path towards
understanding some of the crucial features of our results, and
consider it in Sec.~\ref{sec:CYL}. Sec.~\ref{sec:QCYL} is devoted
to the calculations for more realistic quasi-cylindrical Fermi
surface (Fig.~\ref{fig:model}), and at the end of it we discuss
the results, compare them with the data on CeCoIn$_5$ and organic
$\kappa-(BEDT)-TTF$.

Our Secs.~\ref{sec:CYL}, ~\ref{sec:QCYL}, and ~\ref{sec:CON} are
intended for those readers who are interested only in the overall
physical picture and the behavior of the measured properties; the
figures in Sec.~\ref{sec:QCYL} show the main differences between
the self-consistent and non-self-consistent calculations. Finally,
our conclusions provide a side-by-side comparison of the specific
heat discussed in I and the thermal conductivity results, and
outline implications for future experiments.

\section{\label{sec:QC} Quasiclassical Approach and the equilibrium Green's function}

\subsection{Basic equations and formulation}

We follow Ref.~I in considering the  quasiclassical (integrated
over the quasiparticle band energy) Green's function in a singlet
superconductor in magnetic field
\cite{eil68,lar68,ser83,ale85,AHoughton:1998,esc99}. In the spin
and particle-hole (Nambu) space the matrix propagator depends on
the direction at the Fermi surface (FS), $\hvp$, and the center of
mass coordinate, $\vR$, and is
    \be \whg(\vR, \hvp; \vare) =
    \left( \begin{array}{cc}
    g & i\sigma_2 f \\
    i\sigma_2 \ul{f} & -g
    \end{array} \right) \,.
    \label{eq:dg}
    \ee

We write the quasiclassical equation for the real-energy, $\vare$,
retarded, advanced, and Keldysh propagators, which enables us to
carry out non-equilibrium calculations, see Appendix
\ref{app:Keldysh} below and
Refs.~\onlinecite{ser83,rai95,MJGraf:1996}. The retarded (R) and
advanced (A) functions $\whg=\whg^{R,A}$ satisfy (we take
convention $e<0$)
    \bea
    [(\vare + {e\over c} \vv_f(\hat{\vp}) \vA(\vR) )\, \widehat{\tau}_3
    - \whDelta(\vR, \hat{\vp}) - \whs_{imp}(\vR; \vare),
    \nonumber \\
    \whg(\vR, \hat{\vp}; \vare)] + i\vv_f(\hat{\vp}) \cdot \gradR \;
    \whg(\vR, \hat{\vp}; \vare) = 0 \,, \label{eq:eil}
    \eea
together with the normalization condition
    \be \whg^{R,A}(\vR,
    \hat{\vp}; \vare_m)^2 = -\pi^2 \widehat{1} \,.
    \label{eq:gnorm}
    \ee
Here $\vv_f(\hat{\vp})$ is the Fermi velocity at a point
$\hat{\vp}$ on the FS. The vector potential $\vA(\vR)$ describes
the applied magnetic field, and the self-energy, $\whs$,
(different for the retarded and the advanced components) is due to
impurity scattering. The mean field singlet order parameter,
    \be \whDelta = \left(
    \begin{array}{cc}
    0 & i\sigma_2 \Delta \\
    i\sigma_2 \Delta^* & 0
    \end{array} \right) \,,
    \ee
is self-consistently determined using the Keldysh functions
$f^K$,
    \be
    \Delta(\vR,\hvp) = \int {d\vare \over 4\pi i}\,  \langle \, V(\hvp,
    \hvp') \, f^K(\vR, \hvp'; \vare)\rangle_\sm{FS} \,
    \label{eq:DeltaSC}\,.
    \ee
 In Eq.(\ref{eq:DeltaSC}) we used   a shorthand notation
    \be
    \langle \; \bullet \; \rangle_\sm{FS} = \int d\hvp_\sm{FS} \;
    n_f(\hvp) \;  \bullet \,,
    \ee
where $n_f(\hvp) = N_f(\hvp)/{\cal N}_f$, with $N_f(\hvp)$ the
density of states (DOS) at a point $\hvp$ on the Fermi surface in
the normal state,  and  ${\cal N}_f = \int d\hvp_\sm{FS}
N_f(\hvp)$ the net density of states.

Throughout our work we use a separable pairing interactions,
    \be
    V(\hat{\vp},\hat{\vp}') = V_s \, \cY(\hat{\vp}) \, \cY(\hat{\vp}')
    \,,
    \ee
where $\cY(\hat{\vp})$ is the normalized basis function for the
angular momentum representation, $\langle
\cY(\hat{\vp})^2\rangle_{\sm{FS}} = 1$. Hence the order parameter
is $\Delta(\vR,\hat{\vp}) = \Delta(\vR) \cY(\hat{\vp})$. For
example, for a 2D $d_{x^2-y^2}$ gap we take
$\cY(\hat{\vp})=\sqrt2 ({\widehat p}_x^2-{\widehat p}_y^2)$.

We include the impurity scattering via the self-consistent
$t$-matrix approximation, with the self-energy
    \be \whs_{imp}(\vR; \vare) = \left( \begin{array}{cc}
    D+\Sigma & i\sigma_2 \Delta_{imp} \\
    i\sigma_2 \ul{\Delta}_{imp} & D-\Sigma
    \end{array} \right) = n_{imp} \hat{t}(\vR;\vare) \,.
    \label{eq:sigma-imp}
    \ee
Here $n_{imp}$ is the impurity concentration, and, if $u$ is the
isotropic single impurity potential, the $t$-matrix is defined via
    \be \hat{t}(\vR; \vare)
    = u \hat{1} + u {\cal N}_f \langle \whg(\vR,\hvp;\vare) \rangle_\sm{FS}
    \; \hat{t}(\vR; \vare)\,.
    \label{eq:t-matrix}
    \ee
The component labeled $D\,\widehat{1}$ drops out of the equations
for the retarded and the advanced Green's functions since the unit
matrix commutes with the Green's function in Eq.~(\ref{eq:eil}).
This term, however, appears in the Keldysh part, and affects the
transport properties \cite{PJHirschfeld:1988,lof04} (see also
Appendix~\ref{app:Keldysh}). Below we parameterize the scattering
by the ``bare'' scattering rate, $\Gamma=n_{imp}/\pi {\cal N}_f$,
and the phase shift $\delta_0$ of the impurity scattering, $\tan
\delta_0=\pi u {\cal N}_f$.

\subsection{Equilibrium Green's function}

In I we solved the quasiclassical equations in the vortex state.
We took the superconducting order parameter in the form $\Delta
(\bm R, \hvp)=\Delta(\bm R)\cY(\hvp)$, where
    \begin{subequations}
     \bea
     &&\Delta(\bm R)= \sum_n\Delta_n \braket{\vR}{n}
     \label{eq:Dn}
     \\
     &&\braket{\vR}{n}=\sum_{k_y} C_{k_y}^{(n)} {e^{ik_y\sqrt{S_f} y}
    \over \sqrt[4]{S_f \Lambda^2}} \widetilde\Phi_n\left( x,k_y \right) \,.
    \label{eq:Rn}
    \eea
    \label{eq:VL}
    \end{subequations}
We showed that for $\sum_{k_y} | C_{k_y}^{(n)} |^2=1$ the
coefficients $C_{k_y}^{(n)}$ determine the shape of the lattice,
while $\Delta_n$ is the amplitude of the order parameter in the
$n$-th Landau level channel. The expansion of $\Delta(\bm R)$ is
in the Landau level function of the renormalized coordinates,
\begin{equation}
  \widetilde\Phi_n(x,k_y)=\Phi_n\left( {x-\Lambda^2
    \sqrt{S_f} k_y\over \Lambda \sqrt{S_f}} \right) \,.
    \label{Phi-n}
\end{equation}
The anisotropy factor, $S_f$, for the field applied at an angle
$\theta_H$ to the $z$-axis, is
    \be S_f = \sqrt{ \cos^2 \theta_H +
    {v_{0||}^2\over v_{0\perp}^2} \sin^2 \theta_H}  \,,
    \ee
where  $v_{0\perp}^2 = 2 \langle \cY^2(\hvp) v^2_{\perp i}(p_z)
\rangle_\sm{FS}$ and $v_{0\parallel}^2 = 2 \langle \cY^2(\hvp)
v^2_\parallel(p_z) \rangle_\sm{FS}$; here $v_\parallel$ is the
projection of the Fermi velocity on the $z$ axis, and $v_{\perp
i}$ with $i=x,y$ is the projection on the axes in the plane normal
to $z$.

Following the BPT procedure we replaced the diagonal part of the
Green's function, $g$, with its spatial average, and introduced
the ladder operators for the Landau levels which allowed us to
solve for the anomalous (Gorkov) functions in terms of $g$,
\begin{subequations}
    \be
    f(\vR, \hat{\vp}; \vare) = \sum_m f_m(\hvp, \vare) \braket{\vR}{m} \,,
    \ee
    \be
    f_m(\hvp, \vare) = ig \, \sum_n (-\tilde{v}_-(\hvp))^{m-n} \,
    \cD_{m,n}(\vare, |\hvp|) \widetilde{\Delta}_n(\hvp; \vare) \,.
    \ee
    \label{eq:f-res}
\end{subequations}
Here
    \be \tilde{v}_\pm (\hat{\vp})=
    \frac{\tilde{v}_f(\hat{\vp})_x \pm i \tilde{v}_f(\hat{\vp})_y}
      {|\tilde{v}_f^\perp|} \,,
    \ee
with
    \be
    \tilde{v}_f(\hat{\vp})_x = v_f(\hat{\vp})_x /\sqrt{S_f}  \quad ,\quad
    \tilde{v}_f(\hat{\vp})_y = v_f(\hat{\vp})_y \sqrt{S_f} \,,
    \ee
and
    \be
    |\tilde{v}_f^\perp (\hat{\vp})| =
    \sqrt{\tilde{v}_f(\hat{\vp})_x^2 + \tilde{v}_f(\hat{\vp})_y^2}\,
    \ee
The coefficients of the expansion are given by
        \be
    \cD_{m,n}(\vare, |\hvp|) = \sqrt{\pi}{2 \Lambda \over |\tilde{v}^\perp_f|}
    \sum_{j=0}^{min(m,n)} (-1)^{n_1}
    D_{m,n}^{n_1, n_2} \left({2\tilde{\vare}\Lambda\over |\tilde{v}^\perp_f|} \right) \,,
    \label{eq:Dmn}
    \ee
with $n_1(j)=j+(|m-n|-(m-n))/2$, $n_2(j)=j+(|m-n|+(m-n))/2$ in
each term and
    \be
    D^{n_1,n_2}_{m,n}(z) =
    \left({-i\over\sqrt{2}}\right)^{n_1+n_2}
    {\sqrt{n!} \sqrt{m!} \over (n-n_1)! n_1! n_2!} W^{(n_1+n_2)}(z) \,,
    \ee
where $W^{(n)}(z)$ is the $n$-th derivative of the function $W(z)
= \exp(-z^2) \mbox{erfc}(-iz)$.

We then use the normalization condition,
    \be g^2 -
    \overline{\ul{f} \, f} = -\pi^2 \,,
    \ee
in the spatially averaged form, with $\overline{f_1\, f_2}$ the
spatial average of the product,
to find the equilibrium Green's function
        \begin{subequations}
    \bea
    g &=& -i \pi / \sqrt{1+P}\, ,
    \\
    P &=& -i\sqrt{\pi} {2\over w^2} \sum_n \sum_m \widetilde{\ul\Delta}_n \widetilde{\Delta}_m
    \sum'_{k,l\ge0} \frac{(\tilde{v}_+)^l(-\tilde{v}_-)^k}{l! \, k!}
    \nonumber \\
    &&\hspace{-1cm} \times \bra{n} a^{\dag k} a^l \ket{m} \left( {-i\over \sqrt{2}} \right)^{k+l}
    W^{(k+l+1)}\left( {\sqrt{2}\tilde{\vare}\over w}\right) \,,
    \eea
    \label{eq:g-res}
    \end{subequations}
where $w=|\tilde{v}^\perp_f|/ \sqrt{2}\Lambda$, and the prime over
the $k,l$-sum denotes the restriction that the matrix element
$\bra{n} a^{\dag k} a^l \ket{m} = \sqrt{n!m!/(n-k)!(m-l)!}$ is
non-zero only for $k\le n$, $l\le m$ and $k-l=n-m$. This
expression reduces to the form of $g$ obtained previously if we
truncate the order parameter expansion at the lowest Landau
level~\cite{WPesch:1975,PKlimesch:1978,AHoughton:1998,IVekhter:1999,HKusunose:2004}.
    \be
    g = \frac{-i \pi}{
    \sqrt{1-i\sqrt{\pi}\left(\frac{2\Lambda}{|\tilde{v}_f^\perp|}\right)^2
    \, W^\prime( \frac{2\tilde{\vare}\Lambda}{|\tilde{v}_f^\perp|} ) \,
    \widetilde\Delta_0 \ul{\widetilde\Delta}_0 } } \,.
    \label{eq:g0}
    \ee
This latter form is useful for semi-analytical calculations.

\section{\label{sec:Thermo} Heat conductivity}

\subsection{Linear response and thermal conductivity}

We now proceed to derivation and analysis of expression for the
thermal conductivity in the linear response theory. We first
derive the general formula for the heat conductivity tensor, based
on the closed-form solution for the quasiclassical retarded and
advanced propagators found above, and using the non-equilibrium
Keldysh approach. The Keldysh part of the full quasiclassical
Green's function carries information about both the spectrum and
the distribution of quasiparticles, and the heat current is
defined as energy transfer by
quasiparticles\cite{ser83,MJGraf:1996}
    \be
    \vj_h(\vR) = 2 N_f \int d\hvp \; n_f(\hvp)\vv_f(\hvp)
    \int\limits_{-\infty}^{+\infty} {d\vare\over 4\pi i }
    \vare  g^K(\hvp, \vR; \vare) \,,
    \ee
where $g^K$ is the diagonal component of the Keldysh
propagator.

In equilibrium $\vj_h=0$ as expected, and  in linear response we
define the heat conductivity tensor $\hat{\kappa}$ via $\vj_h =
-\hat{\kappa} \grad T$. We linearize the equations to find the
first order corrections to the retarded, advanced and Keldysh
propagators, $\delta\whg^{R,A,K}$, with respect to $\grad T$. This
implicitly assumes that the inhomogeneity due to the temperature
gradient is much smaller and occurs on much longer scales than the
inhomogeneity due to vortices, impurities, etc., which is the case
experimentally. The details of the derivation of $\kappa$ are
presented in Appendix~\ref{app:Keldysh} and here we give only the
final expression,
\begin{widetext}
\begin{subequations}
    \be
    {\kappa_{ij} \over T} =
    \frac{N_f}{2} \int {d\vare\over T} 
    \frac{\vare^2}{T^2 \cosh^2(\vare/2 T)} 
    \int d\vp_\sm{FS} \, n_f(\hvp) \;
    v_{f,i}(\hat{\vp}) v_{f,j}(\hat{\vp}) \; \frac{-i G_2}{G_1 G_4-G_2 G_3} (-{1\over\pi} \Im g^R )
    \label{eq:hcon}
    \ee
where we defined,
    \bea
    G_1 &=& -D^a + \frac{1}{2(g^R - g^A)} \, \overline{[
    -(f^R + f^A) (\ul{\widetilde{\Delta}}^R - \ul{\widetilde{\Delta}}^A)
    +(\ul{f}^R + \ul{f}^A) (\widetilde{\Delta}^R - \widetilde{\Delta}^A)]} \\
    G_2 &=& -\Sigma^a + \frac{1}{2(g^R - g^A)} \, \overline{[
    (f^R - f^A) (\ul{\widetilde{\Delta}}^R - \ul{\widetilde{\Delta}}^A)
    +(\ul{f}^R - \ul{f}^A) (\widetilde{\Delta}^R - \widetilde{\Delta}^A)]} \\
    G_3 &=& -\Sigma^a + \frac{1}{2(g^R - g^A)} \, \overline{[
    (f^R + f^A) (\ul{\widetilde{\Delta}}^R + \ul{\widetilde{\Delta}}^A)
    +(\ul{f}^R + \ul{f}^A) (\widetilde{\Delta}^R + \widetilde{\Delta}^A)]} \\
    G_4 &=& -D^a + \frac{1}{2(g^R - g^A)} \, \overline{[
    -(f^R - f^A) (\ul{\widetilde{\Delta}}^R + \ul{\widetilde{\Delta}}^A)
    +(\ul{f}^R - \ul{f}^A) (\widetilde{\Delta}^R + \widetilde{\Delta}^A)]}
    \,,
    \eea
\end{subequations}
\end{widetext}
and used the following notations:
$D^a(\vare)=D^R(\vare)-D^A(\vare)$,
$\Sigma^a(\vare)=\Sigma^R(\vare)-\Sigma^A(\vare)$, and
$\widetilde{\Delta}^{R,A}(\vR, \hvp;\vare) = \Delta(\vR,\hvp) +
\Delta^{R,A}_{imp}(\vR;\vare)$. In both Born and unitarity
scattering limits $D^a=0$, which simplifies
this result\cite{PJHirschfeld:1988,MJGraf:1996}.

We can re-write Eq.~(\ref{eq:hcon}) as
\begin{eqnarray}
&&\frac{\kappa_{ij}(T,H)}{T} = \int\limits^{+\infty}_{-\infty} \;
\frac{d\vare}{2 T} \frac{\vare^2}{T^2} \cosh^{-2}\frac{\vare}{2T}
\\
\nonumber &&\times \int \, d\vp_\sm{FS} \, v_{f,i} v_{f,j} \, N(T,
\bm H; \hat{\vp}, \vare) \; \tau_H(T, H; \hat{\vp},\vare) \,.
\end{eqnarray}
Here $N(T, H; \hat{\vp},\vare) = N_f n_f(\hvp) (-\Im \, g^R/\pi)$
is the angle-dependent DOS, and $\tau_H = -i G_2/(G_1 G_4 - G_2
G_3)$ has the meaning of the {\em transport} lifetime due to both
impurity and vortex scattering. In the normal state $\tau_H =
\tau_n = 1/2\gamma$ ($\gamma=\Gamma \sin^2\delta_0$) and $-\Im \,
g^R/\pi=1$. Notice that the transport and the single-particle
lifetimes are different.

Several limiting cases are useful for developing a qualitative
understanding of the physical picture. In the Born or unitary
limit $D^R=D^A=0$. If we truncate the expansion of the vortex
state at the lowest Landau level function, $n=0$, and neglect the
off-diagonal impurity self-energy $\Delta_{imp}=0$, we obtain from
Eqs.(\ref{eq:f-res}-\ref{eq:g-res})
    \bea
    g^R &=& \frac{-i \pi}{
    \sqrt{1-i\sqrt{\pi}\left(\frac{2\Lambda}{|\tilde{v}_f^\perp|}\right)^2
    \, W^\prime( \frac{2\tilde{\vare}\Lambda}{|\tilde{v}_f^\perp|} ) \, |\Delta_0 \cY(\hat\vp)|^2 } } \,
    \label{eq:g00}
    \\
    \overline{\underline{\Delta}f^R} &=& i g^R \; \frac{2\sqrt{\pi} \Lambda}{|\tilde{v}_f^\perp|}
    W\left( \frac{2\tilde{\vare}\Lambda}{|\tilde{v}_f^\perp|} \right) \; |\Delta_0 \cY(\hat\vp) |^2\,.
    \label{eq:f0}
    \eea

In this approximation $\widetilde{\Delta}^R =
\widetilde{\Delta}^A = \Delta$ and thus $G_1=0$, so for the
thermal transport lifetime we find,
    \be
    \frac{1}{2\tau_H} =
    - \Im \Sigma^R + \sqrt{\pi}{2 \Lambda \over |\tilde{v}_f^\perp|}
    \frac{\Im[g^R \, W(2\tilde{\vare}\Lambda/|\tilde{v}_f^\perp|)]}
    {\Im \, g^R} |\Delta_0 \cY|^2
    \,.
    \label{eq:tauH}
    \ee
which agrees with results in Ref.~\onlinecite{IVekhter:1999}. We,
however, aim to include the higher components of the order
parameter expansion for a fully self-consistent calculation and
for comparison with experiment.

\subsection{General properties of the thermal conductivity tensor}


As in I we focus on a tetragonal system with an open, along the
$z$-axis, Fermi surface, and the magnetic field applied in the
$xy$ plane, at angle $\phi_0$ to the $x$-axis. We consider both
$d_{x^2-y^2}$ and $d_{xy}$ order parameters, and model their
variation around the Fermi surface by $\cY(\phi)=\sqrt 2\cos
2\phi$ and $\cY(\phi)=\sqrt 2\sin2\phi$ respectively, where $\phi$
is the angle between the projection of the Fermi momentum on the
basal plane and the $x$ axis. As before, we will consider both a
cylindrical (no energy dispersion along $z$), and
quasi-cylindrical (tight-binding dispersion along $z$) Fermi
surfaces. The following considerations are valid irrespective of
the Fermi surface shape.

\begin{figure}[t]
\centerline{\includegraphics[height=4cm]{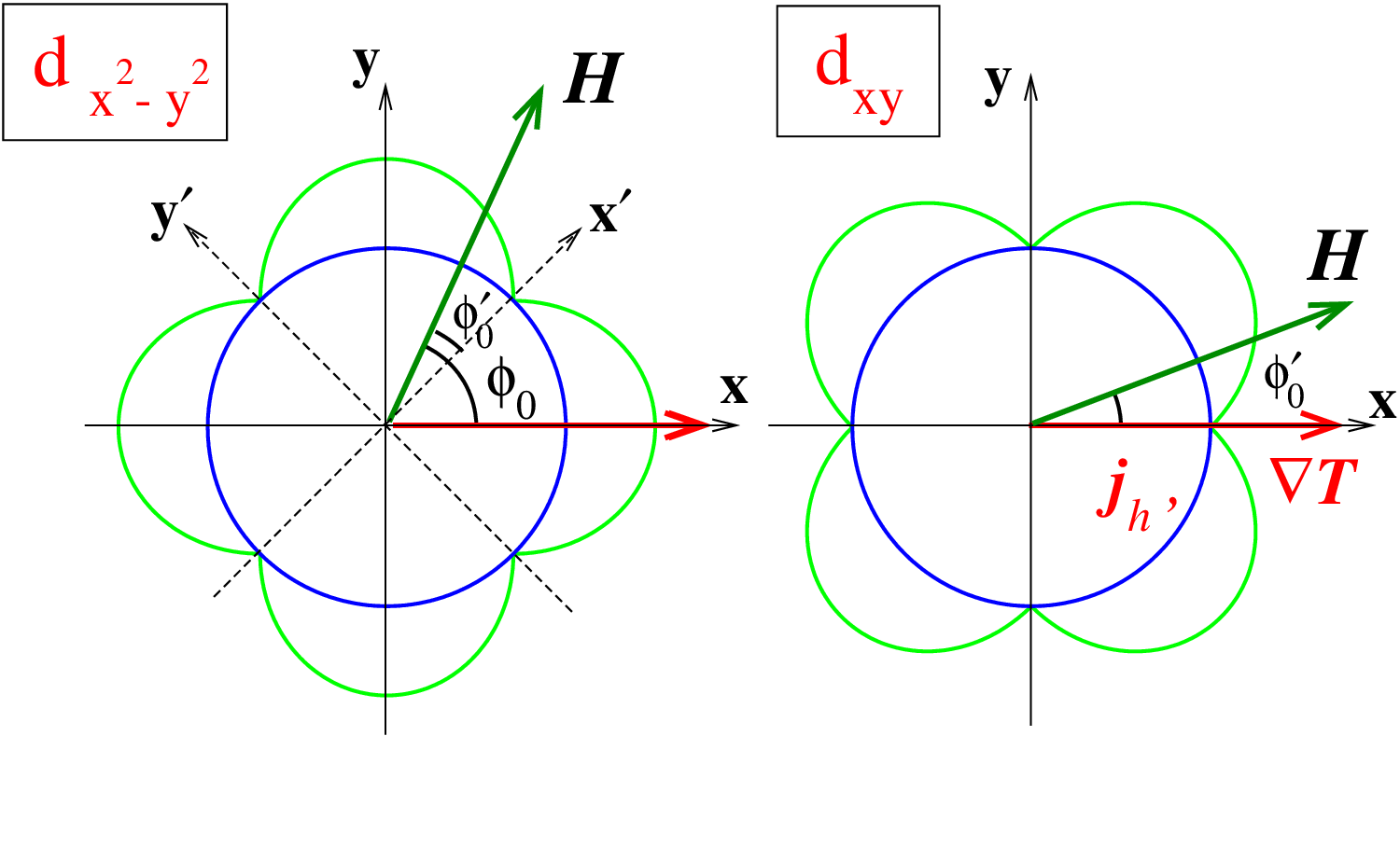}}
\caption{\label{fig:OP-Therm} (Color online) Two distinct
experimentally relevant setups for the thermal conductivity. Left
panel: $d_{x^2-y^2}$ gap symmetry, right panel: $d_{xy}$ gap
symmetry. Response of the $d_{xy}$ superconductor to the thermal
gradient along the [100] ($\vx$) direction with the field at an
angle $\phi_0^\prime$ to this axis is equivalent to the response
of a $d_{x^2-y^2}$ system to the thermal gradient along the [110]
($\vx^\prime$) direction and the field at the angle
$\phi_0=\phi^\prime+\pi/4$ to the $\vx$-axis. Note that the
expriment is done with the heat current $\vj_h$ along [100], while
the calculations are for the thermal gradient along this
direction, see text for details.}
\end{figure}

Experimentally, the in-plane (interplane) heat conductivity is
measured by driving the heat current along a high symmetry
crystalline direction, such as [100] or [110] ([001]). The
longitudinal and/or transverse thermal gradient are defined and
measured with respect to the direction of the heat current. This
creates two physically distinct cases for the in-plane transport:
the heat flow in the experiment is along either a node or
antinode, see Fig.~\ref{fig:OP-Therm}. If our task is, for
example, to determine the shape of the gap from the measured
thermal conductivity along the $x$-axis we cannot {\em a priori}
assume whether the heat current is along a node or a gap maximum.

From the theoretical perspective, the knowledge of the full
thermal conductivity tensor, Eq.~(\ref{eq:hcon}) allows to
determine the heat transport along an arbitrary direction. The two
cases, $d_{x^2-y^2}$ and $d_{xy}$, see Fig.~\ref{fig:OP-Therm},
transform into one another by rotation of the heat current:
thermal conductivity measured for the heat current in the [100]
direction for the $d_{xy}$ gap is equal to the thermal
conductivity for the $d_{x^2-y^2}$ order parameter with the heat
current in [110] direction.  Therefore we focus on the
$d_{x^2-y^2}$, and compute all the components $\kappa_{ij}$ in the
plane; the thermal conductivity for the $d_{xy}$ case is computed
using these results.  In a tetragonal system in the absence of the
field, the off-diagonal elements $\kappa_{xy}=\kappa_{yx}=0$, and
the diagonal elements are equal, $\kappa_{xx}=\kappa_{yy}$, so
that the conductivity is isotropic. Applying a magnetic field in
the plane changes the situation dramatically. First, for the field
applied at an angle $\phi_0$ relative to the [100] direction,
$\kappa_{xx}\neq\kappa_{yy}$;  it is easy to see (and we show it
formally below) that
$\kappa_{yy}(\phi_0)=\kappa_{xx}(\pi/2+\phi_0)$ since for these
components the angle between the field and the heat current is the
same.

Second, the in-plane field breaks the tetragonal symmetry, and
therefore $\kappa_{xy}(\phi_0)\neq 0$ for a general orientation of
the field. We emphasize that this occurs even when the Lorenz
force is neglected. The non-vanishing $\kappa_{xy}$ arises because
of the difference between the transport along the vortex and
normal to it; when the field (and the vortices) are at an
arbitrary angle to the direction of the heat current, a transverse
temperature gradient appears similar to the Hall conductivity in a
material with the electric field applied at an arbitrary angle to
inequivalent principal axes. The transverse heat conductivity is
of the same order or magnitude as the anisotropy between transport
parallel and normal to the vortices, and hence much greater than
the typical contribution proportional to the cyclotron frequency,
$\omega_c=eH/mc\sim (\Delta^2/E_f)(H/H_{c2})$. Therefore the Hall
angle is moderately large, see below.

To make the argument more rigorous consider the general form of
the heat conductivity tensor as a function of the field
orientation, $\hat{\vH}$. We write Eq.~(\ref{eq:hcon}) in the
form,
    \be
    \kappa_{ij}(\phi_0) =
    \langle
    v_{f,i}(\hvp) v_{f,j}(\hvp) \; \cK((\hvp_f\cdot\hat{\vH})^2, |\Delta(\hvp)|^2; T, H)
    \rangle \,.
    \label{eq:hconkern}
    \ee
Here the kernel $\cK((\hvp_f\cdot\hat{\vH})^2, |\Delta(\hvp)|; T,
H)$ is determined by the equilibrium Green's functions, and, at a
given point ${\widehat{\vp}}_f$ at the Fermi surface, depends on
the angle between the Fermi velocity and the field,
$(\hvp_f\cdot\hat{\vH})$, the gap amplitude for that direction,
$|\Delta(\hvp)|$, as well as on $T$ and $H$.  Since the kernel
does not change if the direction of the field is reversed, we
explicitly write it as dependent on $(\hvp_f\cdot\hat{\vH})^2$

Let us start by considering a $d_{x^2-y^2}$ gap. The inversion of
the field in the $xy$-plane corresponds to the change $\phi_0 \to
\phi_0 + \pi$. We can simultaneously change the variables in
Eq.~(\ref{eq:hconkern}) according to $(\hp_x, \hp_y) \to (-\hp_x,
-\hp_y)$, which leaves the kernel invariant, and find
    \be
    \kappa_{ij}(\phi_0+\pi) = \kappa_{ij}(\phi_0) \,,
    \label{eq:symKinv}
    \ee
for $i,j = \{x,y\}$; at the same time $\kappa_{iz}(\phi_0+\pi) =
-\kappa_{iz}(\phi_0)$, and $\kappa_{zz}(\phi_0+\pi) =
\kappa_{zz}(\phi_0)$. Similarly, reflection of the field in the
$xz$-plane, $\phi_0 \to -\phi_0$, together with reflection $\hp_y
\to -\hp_y$ again does not change the kernel, and leads to
    \bea
    \kappa_{xx}(-\phi_0) &=& \kappa_{xx}(\phi_0) \,,
    \nonumber \\
    \kappa_{yy}(-\phi_0) &=& \kappa_{yy}(\phi_0) \,,
    \label{eq:symKevenodd}
    \\
    \kappa_{xy}(-\phi_0) &=& -\kappa_{xy}(\phi_0) \,.
    \nonumber
    \eea
Finally, if we rotate the field and the coordinate system, $\phi_0
\to \phi_0 + \pi/2$ [$(\hp_x, \hp_y) \to (\hp_y, -\hp_x)$], we
find
    \bea
    \kappa_{xx}(\phi_0+\pi/2) &=& \kappa_{yy}(\phi_0) \,,
    \nonumber \\
    \kappa_{yy}(\phi_0+\pi/2) &=& \kappa_{xx}(\phi_0) \,,
    \label{eq:symKrot}
    \\
    \kappa_{xy}(\phi_0+\pi/2) &=& -\kappa_{xy}(\phi_0) \,.
    \nonumber
    \eea

We carry out Fourier expansion based on these symmetries. From
Eq.~(\ref{eq:symKinv}), Eq.~(\ref{eq:symKevenodd}), and
Eq.~(\ref{eq:symKrot}) we find for the $d_{x^2-y^2}$ gap
    \bea
    \kappa_{xx}(\phi_0) &=& \kappa_0 + \kappa_{2} \cos2\phi_0  +
            \kappa_{4} \cos4\phi_0 + \dots \,,
    \nonumber \\
    \kappa_{yy}(\phi_0) &=& \kappa_0 - {\kappa}_{2} \cos2\phi_0  +
            {\kappa}_{4} \cos4\phi_0 + \dots \,,
    \label{eq:Fourier_x2y2}
    \\
    \kappa_{xy}(\phi_0) &=& \tilde{\kappa}_{2} \sin2\phi_0  +
            \tilde{\kappa}_{6} \sin6\phi_0 + \dots \,.
    \nonumber
    \eea
The $\cos 2\phi_0$ term in the longitudinal conductivity describes
the anisotropy between the transport along and normal to the
vortices. Furthermore, if the superconducting gap is isotropic (or
absent), and hence the only dependence of the kernel $K$ on the
field orientation is via the term $(\hvp_f\cdot\hat{\vH})^2$, it
immediately follows that for our cylindrically symmetric Fermi
surface $\kappa_{xx}+\kappa_{yy}$ is independent of the field
orientation, $\phi_0$, which requires $\kappa_{4}=0$. For an
anisotropic Fermi surface there may be an angular modulation of
the thermal conductivity, but it would occur already in the normal
state. Therefore the $\cos 4\phi_0$ component that appears only in
the superconducting state is predominantly due  to the gap
anisotropy. Such a decomposition in the analysis of the
experimentally measured thermal conductivity was used in
Refs.~\onlinecite{KIzawa:CeCoIn5,KIzawa:BEDT,YMatsuda:2006} to
infer the gap structure of heavy fermion and organic
quasi-two-dimensional materials; and in the following section we
compare our results with their analysis.

The origin of the $\sin 2\phi_0$ directional dependence of the
transverse thermal conductivity is also transparent. In the
presence of the field, the principal axes of the thermal
conductivity tensor are along and normal to $\vH$. Consequently,
when the heat current is along one of those axes, no transverse
signal is generated, irrespective of the nodal structure.  This is
precisely the result found in high-T$_c$ superconductors by
Oca\~na and Esquinazi \cite{ROcana:2001,ROcana:2002}, who observed
a nearly perfect sinusoidal thermal Hall response.

We are now in the position to consider the differences between the
$d_{x^2-y^2}$ and $d_{xy}$ gaps. The longitudinal and transverse
conductivities for the $d_{xy}$ order parameter are identical to
the components of the thermal conductivity tensor for the
$d_{x^2-y^2}$ case in the coordinates $(x',y')$ rotated by
$\alpha=\pi/4$ with respect to $(x,y)$, see
Fig.~\ref{fig:OP-Therm},
    \be
    \hat{\kappa}^\prime=\hat{\kappa}_{d_{xy}} = \left( \begin{array}{cc}
    \kappa_{x'x'}  & \kappa_{x'y'}  \\
    \kappa_{x'y'}  & \kappa_{y'y'}
    \end{array} \right) \,.
    \label{eq:kappatensor}
    \ee
Moreover, the field applied at an angle $\phi_0^\prime$ to the
${\bf x}^\prime$ axis makes angle $\phi_0=\phi^\prime+\pi/4$ with
the ${\bf x}$ axis, so that $\hat{\kappa}'(\phi_0^\prime) =
\hat{\cR}(\alpha) \hat{\kappa}(\phi_0^\prime+\pi/4)
\hat{\cR}^{-1}(\alpha)$, with the rotation matrix
    \be
    \hat{\cR} = \left( \begin{array}{cc}
    \cos\alpha & \sin\alpha \\
    -\sin\alpha & \cos\alpha
    \end{array} \right) \,.
    \ee
This leads to
\begin{eqnarray}
        \kappa_{x'x'}(\phi_0^\prime)&=&
    \kappa_0 + \tilde{\kappa}_{2} \cos 2\phi_0^\prime
    - \kappa_{4} \cos4\phi_0^\prime + \dots \,,
        \nonumber \\
        \kappa_{y'y'}(\phi_0^\prime) &=&
    \kappa_0 - \tilde{\kappa}_{2} \cos 2\phi_0^\prime
    - \kappa_{4} \cos4\phi_0^\prime + \dots \,,
    \label{eq:Fourier_xy_primed}
        \\
        \kappa_{x'y'}(\phi_0^\prime) &=& -{\kappa}_{2} \sin 2\phi_0^\prime  + \dots \,.
    \nonumber
    \end{eqnarray}
Importantly, this result implies that the four-fold term in the
longitudinal thermal conductivity depends only on the orientation
of the field with respect to the nodes of the gap. Indeed, let us
restore the dependence on the angle $\phi_0$ measured to the gap
maximum, then
    \begin{eqnarray}
      \kappa_{x'x'}(\phi_0)&=& \kappa_0 + \tilde{\kappa}_{2} \sin 2\phi_0  +
            \kappa_{4} \cos4\phi_0 + \dots \,,
     \\
    \kappa_{x'y'}(\phi_0) &=& {\kappa}_{2} \cos 2\phi_0  +
             \dots \,.
    \label{eq:Fourier_xy}
    \end{eqnarray}
The last term in the longitudinal thermal conductivity for the
$d_{xy}$ order is identical to that for the $d_{x^2-y^2}$ gap. In
other words, {\em independently of the gap symmetry, the fourfold
term in the longitudinal thermal conductivity  simply depends on
the angle between the direction of the in-plane field, and the
antinodal direction of the gap.} Consequently, in the following
sections we will focus both on the overall features of the thermal
transport and specifically on that term.

\subsection{Calculated vs. measured thermal conductivity}
\label{sec:calc_vs_measur}

We will see below that the field-induced anisotropy in the
transport along and normal to the vortices leads to the large
thermal Hall angle, $\kappa_{xy}/\kappa_{xx}\sim 0.1$. In this
case it is important to keep in mind that theoretical calculations
are done under assumptions different from the typical steady-state
experimental setup. The thermal conductivity tensor is defined via
$\vj_{h,i}=-\kappa_{ij} \grad_j T$, where $\vj_h$ is the
heat current. The experiments are done by driving the thermal
current along a given ([100]) axis, while thermally insulating the
sample in the transverse direction. The experiment measures the
thermal gradients established under the conditions $\vj_{h,x}=j_h$
and $\vj_{h,y}=0$. Consequently, the measured longitudinal,
$\kappa_l$ and the transverse, $\kappa_t$, thermal conductivities
are
\begin{eqnarray}
  \kappa_l&=&\kappa_{xx}-\frac{\kappa_{xy}^2}{\kappa_{yy}}\, ,
  \\
  \kappa_t&=&\frac{{\bm \nabla}_y T}{{\bm \nabla}_x T}\kappa_l
  \approx\kappa_{xy}\frac{\kappa_{xx}}{\kappa_{yy}}\, .
\end{eqnarray}
Presence of the off-diagonal terms does not substantially modify
the absolute value of the longitudinal or transverse conductivity
since $\kappa_{xy}^2/(\kappa_{yy}\kappa_{xx})\sim 0.01$ at most,
and therefore $\kappa_l\approx\kappa_{xx}$ and
$\kappa_t\approx\kappa_{xy}$.

 Note, however, that our principal interest is in the
fourfold nodal term, $\kappa_4$, which is itself only a fraction
of the longitudinal thermal conductivity. Assuming
$\kappa_0\gg\kappa_2, \kappa_4$ we find
\begin{equation}
  \kappa_l^{(4)}\approx
  \kappa_4+\frac{\tilde{\kappa}_2^2}{2\kappa_0}.
  \label{Expt_kappa_4fold}
\end{equation}
In some region of the phase diagram, where $\kappa_2\gg \kappa_4$,
the two terms may be comparable. Our results indicate this range
to be rather small. We find  that the magnitude of the fourfold
term is slightly changed by accounting for the difference between
the computed and the measured quantity; however, the main features
remain unmodified. Hence in the following we discuss the overall
features of the thermal conductivity profiles, and only briefly
return to the difference between the computed and measured
anisotropy in the conclusions.

\section{Cylindrical Fermi surface}
\label{sec:CYL}

Once again we begin by considering the anisotropy of the
longitudinal heat conductivity, $\kappa_{xx}(\phi_0)$, for a
cylindrical Fermi surface, ${\bm v}_f=(v_f\cos\phi, v_f\sin\phi,
0)$ for $0\leq\phi\leq 2\pi$ and $-\pi/c\leq k_z\leq \pi/c$, where
$c$ is the $c$-axis lattice spacing. As described in Ref.~I, this
FS does not allow for the self-consistent calculation of the order
parameter in the vortex state for the in-plane field. As before,
we restrict ourselves to the lowest order Landau wave function for
the order parameter, take
$\Delta(T,H)=\Delta(T,0)[1-H/H_{c2}]^{1/2}$, and use the
corresponding results for the thermal conductivity,
Eqns.(\ref{eq:g00}),(\ref{eq:f0}), and (\ref{eq:tauH}). We choose
$H_{c2}^{ab}$ to be direction independent, and carry out the
self-consistent calculation in temperature and impurity scattering
according to Eqs.(\ref{eq:g00}) and (\ref{eq:tauH}). The impurity
self energy is determined in the unitarity limit with the normal
state mean free path $\ell_n/\xi_0=70$, where $\xi_0$ is the
coherence length.  This toy model lends itself easily to numerical
and, in some limits, semi-analytical work, and therefore allows
investigation of the salient features of the behavior of
$\kappa_{xx}$. We show below that this model gives qualitatively
correct results for quasi-two dimensional systems. In the
self-consistent calculation there is a node-antinode anisotropy in
the upper critical field at low $T$, and the comparison (given
below) between the cylindrical and corrugated Fermi surfaces
elucidates the role of this anisotropy for the behavior of the
thermal conductivity.

In general, to determine $\kappa_{xx}$ we need to
self-consistently determine the DOS and the single particle
lifetime as in the calculation of the specific heat, and then
determine the transport time and the thermal conductivity. For the
lowest Landau approximation these are given by
Eqs.~(\ref{eq:g00}-\ref{eq:tauH}). At finite energies this
procedure can only be carried out numerically, as shown below.
First we make some analytical estimates at low temperatures,
$T=0$, and therefore set $\vare=0$. We consider a $d_{x^2-y^2}$
gap and focus on three values of the thermal conductivity: along
the field, $\kappa_{xx}(\phi_0=0^\circ)$, normal to the field,
$\kappa_{xx}(\phi_0=90^\circ)$, and for the field along the node,
$\kappa_{xx}(\phi_0=45^\circ)$.

Define the mean free path $\ell=v_f\tau_0$, where
$(2\tau_0)^{-1}=- \Im \Sigma^R (\vare=0)$ is the single particle
lifetime, which depends on the net density of states, and hence is
sensitive to the direction of the field, $\phi_0$, via the
self-consistent $T$-matrix equation. The argument of the
$W$-function and its derivatives in
Eqs.~(\ref{eq:g00}-\ref{eq:tauH}) is then $z=i\Lambda/\ell_\perp$,
where $\ell_\perp=\ell|\sin(\phi-\phi_0)|$, and depends on the
position, $\phi$, on the Fermi surface.  Since we work in the
regime $\ell=v_f\tau_0\gg\Lambda$, we can set $z=0$ for most
values of $\phi$, except for the directions nearly parallel to the
field, $|\phi-\phi_0|\leq \Lambda/\ell$ and $|\phi-\phi_0-\pi|\leq
\Lambda/\ell$.  Let us denote the contribution from this narrow
range as $\kappa_1$, and the contribution from the angles outside
of this range as $\kappa_2$, so that
$\kappa_{xx}=\kappa_1+\kappa_2$.

We now estimate each contribution. For $|\phi-\phi_0|\leq
\Lambda/\ell$ we use the expansion for large argument,
$W(z)\approx i/(\sqrt\pi z)$ and $W^\prime(z)\approx -i/(\sqrt\pi
z^2)$ to estimate the contribution to the thermal conductivity as
\begin{eqnarray}
  \frac{\kappa_1(\phi_0)}{TN_f\, v_f^2}&\approx& \frac{4\pi^2}{3} \tau_0
    (\phi_0) \int_{\phi_0-\Lambda/\ell}^{\phi_0+\Lambda/\ell}
  \frac{\cos^2\phi\, d\phi}{[1+(\Delta\tau_0)^2]^{3/2}}
  \\
  &\simeq&
  \frac{4\pi^2}{3} \tau_0
    (\phi_0)\frac{a_1(\phi_0)}{[1+(\Delta\tau_0)^2]^{3/2}},
\end{eqnarray}
where $a_1(0^\circ)=2\Lambda/\ell$, $a_1(45^\circ)=\Lambda/\ell$,
and $a_1(90^\circ)=2\Lambda^3/3\ell^3$.

Over the remainder of the Fermi surface we set $z=0$ and use
$W(0)=1, W^\prime(0)=2i/\sqrt\pi$ to find for $T\rightarrow 0$
\begin{eqnarray}
    \frac{\kappa_2(\phi_0)}{TN_f\, v_f^2}&=&\frac{2\pi^2}{3} \tau_0
    (\phi_0)
    \\
    \nonumber
    &\times&
    \int^\prime \frac{d\phi}{2\pi}\,
    \frac{\cos^2\phi}{\sqrt{1+\delta^2\frac{\cos^2 2\phi}{\sin^2(\phi-\phi_0)}}}
    \frac{1}{1+\mu \frac{\cos^2 2\phi}{|\sin(\phi-\phi_0 )|}}\, ,
    \\
    \delta^2&=& 8(\Lambda\Delta/v_f)^2 \, ,
    \\
    \mu&=&\frac{4\sqrt\pi\Lambda
    \Delta^2\tau}{v_f}=\frac{\sqrt\pi}{2}\frac{\ell}{\Lambda}
    \delta^2\, .
\end{eqnarray}
Here the prime denotes that we are integrating over the entire
Fermi surface excluding the regions close to the field direction
considered above. Notice that $\mu\gg\delta^2$, and therefore the
behavior of the thermal conductivity is controlled to much greater
extent by the transport lifetime than by the density of states.
The transport lifetime is peaked along the nodal directions.

These observations enable some analytical progress starting from
the high fields, $H\lesssim H_{c2}$. In that case
$\delta^2\ll\mu\ll 1$, and we approximate the density of states by
its normal state value, $N_f$. Consequently, $\tau_0=\tau_n$,
where $\tau_n$ is the normal state scattering rate which has no
dependence on the direction of the magnetic field. Defining
$\kappa_n=\pi^2 N_fv_f^2\tau_n T/3$ we find angular variation of
the thermal conductivity is approximately given by
\begin{equation}
  \frac{\kappa_{xx}(\phi_0)}{\kappa_n}=\int^\prime\frac{d\phi}{2\pi}\,
  \frac{2\cos^2\phi}{1+\mu \frac{\cos^2 2\phi}{|\sin(\phi-\phi_0 )|}}\, ,
  \label{eq:ApproxKapCyl}
\end{equation}
We assume here that $\mu\gg\Lambda/\ell$, which is satisfied
nearly everywhere up to $H_{c2}$ for clean systems. Now consider
the behavior of the thermal conductivity for different directions
of the field just below the upper critical field. For small $\mu$
we find the conductivity along the field,
$\kappa_{xx}(0^\circ)\approx \kappa_n[1+(4/\pi)\mu\ln\mu]$, while
the conductivity normal to the field is
$\kappa_{xx}(90^\circ)\approx\kappa_n(1-28\mu/15\pi)$. Finally,
for the field along a node,
$\kappa_{xx}(45^\circ)\approx\kappa_n(1-16\mu/3\pi)$. Therefore in
the immediate vicinity of the transition at low temperature we
expect
$\kappa_{xx}(90^\circ)>\kappa_{xx}(45^\circ)>\kappa_{xx}(0^\circ)$,
or nearly twofold profile of the thermal conductivity.

At lower fields and $T=0$ we enter the regime $\delta^2\lesssim
1\ll\mu$, where we can still approximate the density of states by
the normal state value, but the thermal transport is restricted by
sharp peaks in the lifetime for nodal quasiparticles, see
Eq.~(\ref{eq:ApproxKapCyl}). Linearizing the gap around the nodal
points and carrying out the integration, we find
$\kappa_{xx}(0^\circ)\approx\kappa_{xx}(90^\circ)\simeq\kappa_n/(2^{1/4}\mu^{1/2})$,
and $\kappa_{xx}(45^\circ)\approx \kappa_n/(2\mu^{1/2})$.
Consequently in this regime we find
$\kappa_{xx}(90^\circ)\geq\kappa_{xx}(0^\circ)>\kappa_{xx}(45^\circ)$,
suggesting a weak minimum for the field along the node.
Remarkably, at low $T$ the conductivity normal to the vortex is
always higher than that parallel to the field, but the amplitude
of this anisotropy, and the relative position of the value of the
thermal conductivity for the field applied along a node both
change between $\mu\ll 1$ and $\mu\gg 1$.

\begin{figure}[t]
\centerline{\includegraphics[height=5.5cm]{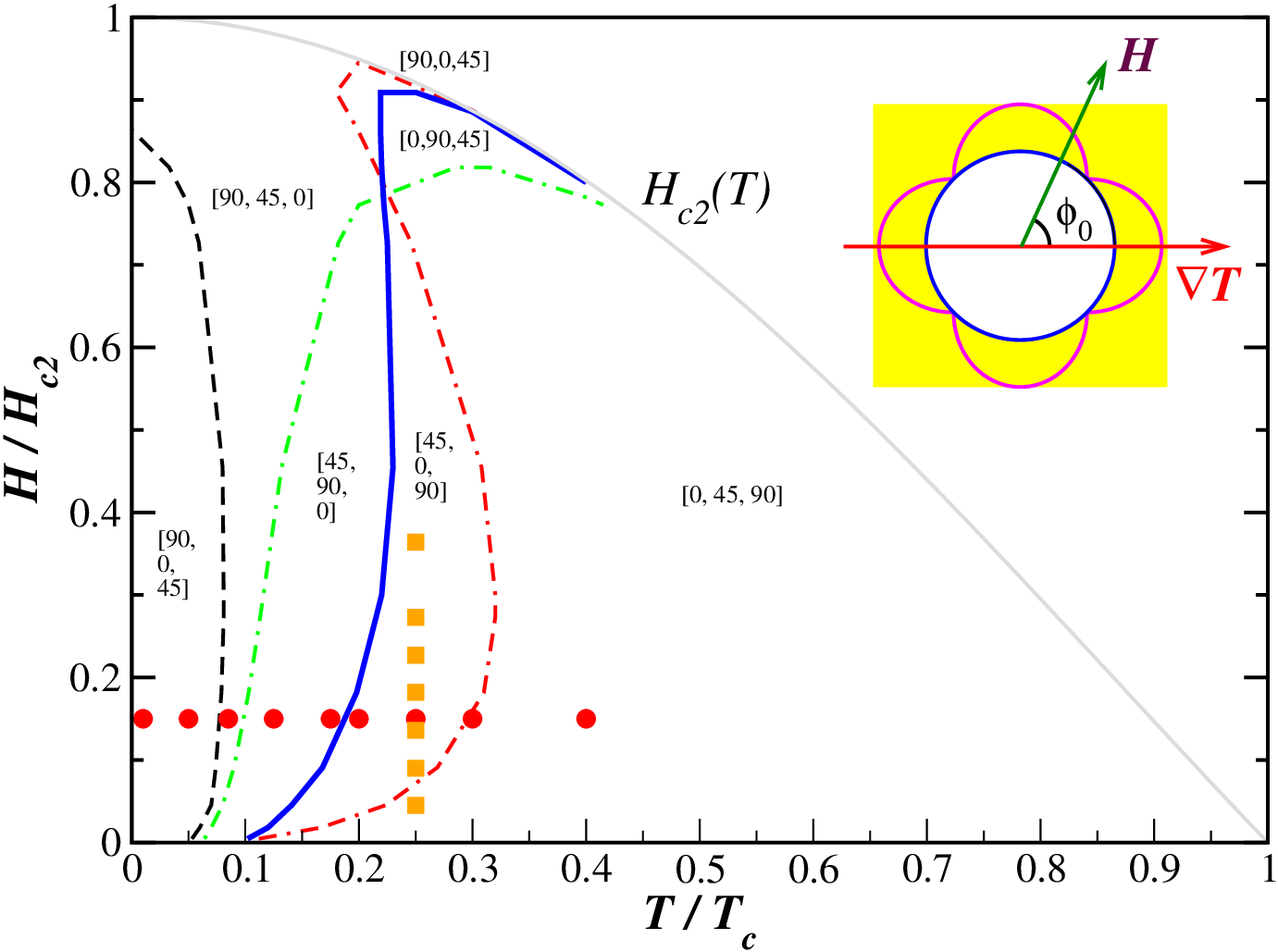}}
\caption{\label{fig:PDcylK} (Color online) Left panel: the
$TH$-diagram of the longitudinal heat conductivity anisotropy,
$\kappa_{xx}(\phi_0)$. The thermal gradient is  along the $x$-axis
(maximal gap). Regions of different anisotropy are marked as
$[\phi_0^a, \phi_0^b, \phi_0^c ]$, which denotes
$\kappa_{xx}(\phi_0^a)
> \kappa_{xx}(\phi_0^b) > \kappa_{xx}(\phi_0^c)$. For points
marked by circles and squares the profiles of the angle-dependent
$\kappa_{xx}(\phi_0)$ are shown in Fig.~\ref{fig:cylPD-Kprof}.}
\end{figure}

\begin{figure}[b]
\centerline{\includegraphics[height=5.5cm]{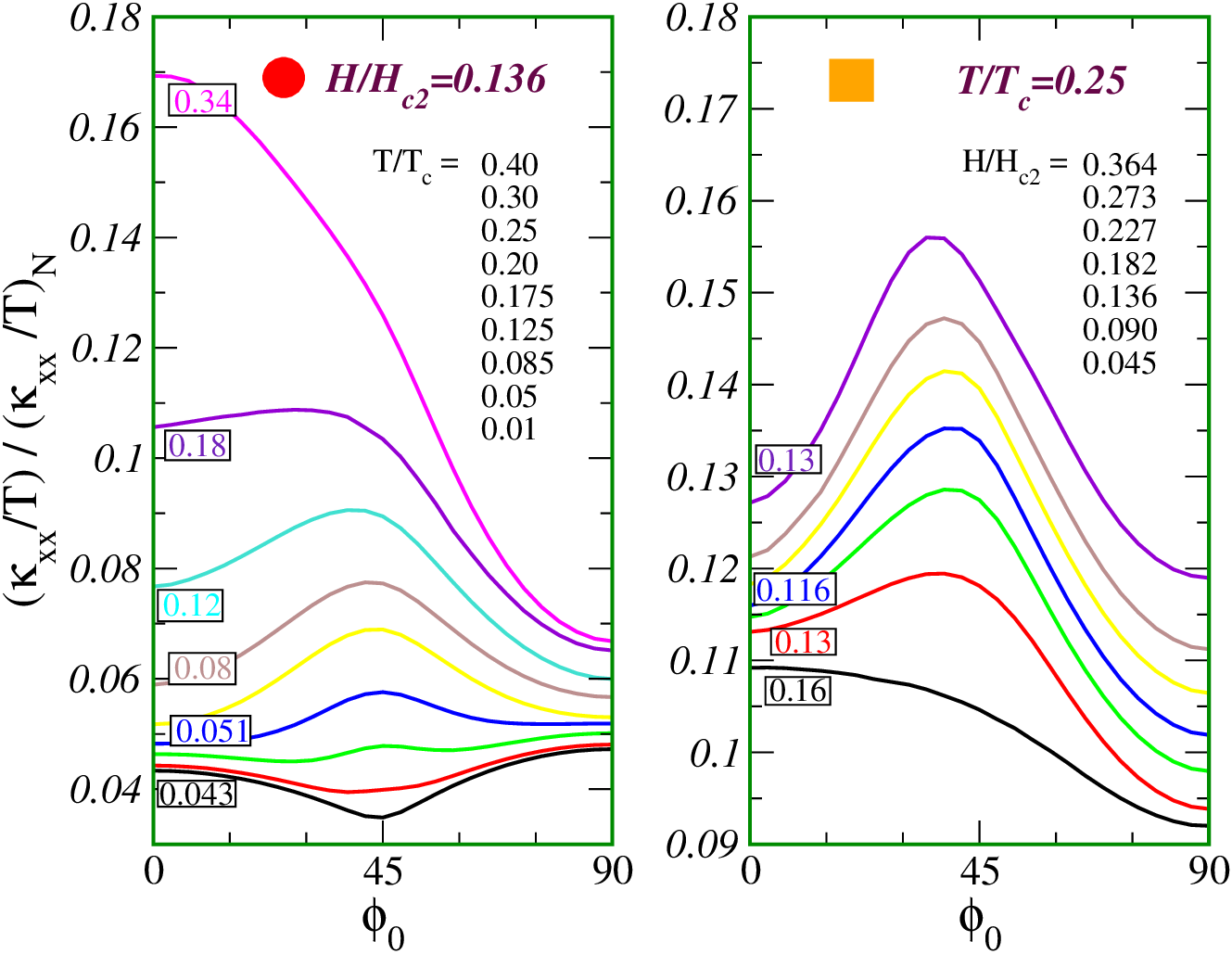}}
\caption{\label{fig:cylPD-Kprof} (Color online) Center panel:
evolution of the heat conductivity with temperature for $H/H_{c2}
\approx 0.14$. Right panel: heat conductivity as a function of the
field angle for different fields at $T/T_c =0.25$. The curves are
shifted vertically for clarity. The framed numbers give true
values of normalized $\kappa_{xx}(\phi_0=0)$ for several of them.
Corresponding temperatures and fields are shown in each panel in
the same order as the curves. }
\end{figure}

This analysis is supported by the numerical results. We show
results only for the longitudinal thermal conductivity,
$d_{x^2-y^2}$ gap, and the heat current along the antinodal
direction. The complete heat conductivity tensor is given and
discussed below for the corrugated Fermi surface, with
self-consistent calculations; The results for both FS are very
similar.

The phase diagram of Fig.~\ref{fig:PDcylK} shows regions with
different anisotropy of the heat conductivity. The changes along
the vertical axis, $T=0$, as a function of the field are in
agreement with our estimates above: at $H\geq 0.85 H_{c2}$ we find
$\kappa_{xx}(90^\circ)>\kappa_{xx}(45^\circ)>\kappa_{xx}(0^\circ)$,
while below that field
$\kappa_{xx}(90^\circ)>\kappa_{xx}(0^\circ)>\kappa_{xx}(45^\circ)$.
Note that for $H/H_{c2}=0.85$ we have $\delta^2\sim 1$ and
therefore already $\mu\gg 1$. The variation of the thermal
conductivity with the the direction of the applied field with
respect to the {\bf x}-axis (angle $\phi_0$) are shown in
Fig.~\ref{fig:cylPD-Kprof} for the points in the $T$-$H$ plane
marked by circles and squares in Fig.~\ref{fig:PDcylK}. Evolution
of $\kappa_{xx}$ with temperature at low fields (circles) is
considerable: minimum for the nodal direction, $\phi_0=45^\circ$,
at low $T$ quickly evolves into a maximum, and at $T/T_c \gtrsim
0.4$ the conductivity is largely twofold with no clear signature
of the nodal structure of the gap. The change of the anisotropy
with the field at moderate $T$ is more gradual (squares), and  a
pronounced peak at $\phi_0=45^\circ$, for the field along the
nodes, persists to moderately high fields, providing a clear
signature of the nodal structure.

The anisotropy between thermal transport normal to and parallel to
the vortices is reversed at moderate temperatures (solid blue line
in Fig.~\ref{fig:PDcylK}): at low temperature $\kappa_{xx}(0) <
\kappa_{xx}(90)$ (in the notation of Eq.~(\ref{eq:Fourier_x2y2})
this means $\kappa_2<0$), while at high $T$ we find
$\kappa_{xx}(0)
> \kappa_{xx}(90)$ (or $\kappa_2>0$).   This evolution is in agreement with that
for a conventional superconductor,  found by Maki
\cite{KMaki:1967}. Note also that the four-fold term
$\sim\kappa_{4}$ in Eq.~(\ref{eq:Fourier_x2y2}) is most pronounced
at intermediate to low $T$ and $H$.

\section{Quasi-cylindrical Fermi surface}
\label{sec:QCYL}

\subsection{Main results}
\label{sec:QCYL-res}

To solve the  quasiclassical equations self-consistently for the
order parameter, we need a model which allows for the $c$-axis
superconducting currents when the field is applied in the $a$-$b$
plane. Hence we analyze a corrugated quasi-2D Fermi surface given
by
$$p_f^2 = p_x^2 + p_y^2 - (r^2 \, p_f^2) \cos (2 s\, p_z/r^2 p_f) \,,$$
so that the quasiparticle velocity has a nonvanishing
$z$-component, This Fermi surface, with $s=r=0.5$, was considered
in I for the analysis of the specific heat, and we take the same
values of parameters to directly compare the anisotropy of the
heat capacity with that of thermal conductivity. Note that for
this choice the DOS anisotropy in the normal state is $n_f(\hvp) =
1$, and the normal state conductivity anisotropy is
$\kappa^N_{zz}/\kappa^N_{xx}=s^2=0.25$. For this anisotropy the
vortex lattice is still  Abrikosov-like.

For the self-consistent calculation of the order parameter and
$H_{c2}$ we limit ourselves to three Landau level components in
Eq.(\ref{eq:Dn}), $\Delta_0, \Delta_2, \Delta_4$, which is
sufficient for the convergence of the calculation to high
precision. As in I, we take impurity scattering in the unitarity
limit with the strength in the normal state $\Gamma/2\pi T_c =
0.007$ (suppression of the transition temperature to $T_c/T_{c0}
\sim 0.95$, and the mean free path $\ell_{n}/\xi_0 \simeq 70$). We
showed in I that this choice gives the following values of the
critical fields at $T=0$: $H_{c2}^{antinode} \approx 1.45 B_0$,
$H_{c2}^{node} \approx 1.27 B_0$ and $H_{c2}^{c} \approx 0.57
B_0$, where $B_0 = \Phi_0 / 2\pi \xi_0^2$ where $\Phi_0=hc/2|e|$.
For the in-plane anisotropy we have then
$(H_{c2}^{antinode}-H_{c2}^{node})/H_{c2}^{antinode} \sim 15\%$,
and the ratio between the $c$-axis and antinodal directions is
$H_{c2}^{c}/H_{c2}^{antinode}=0.4$, similar to that observed in
CeCoIn$_5$ experimentally.

\begin{figure}[t]
\centerline{\includegraphics[height=4cm]{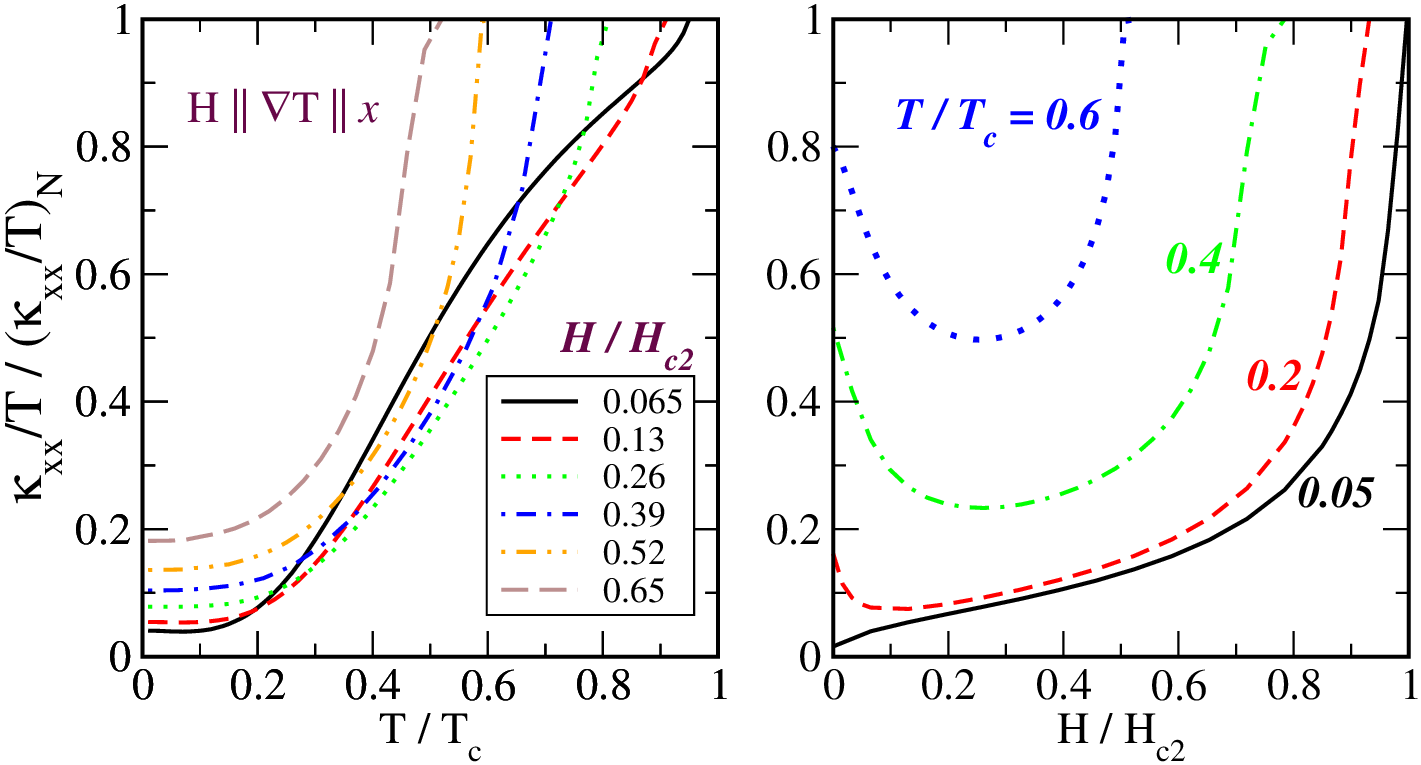}}
\caption{\label{fig:Kbt} (Color online)  Longitudinal heat conductivity for
$\vH || \grad T || \hat{x} $-antinodal direction:
as a function of temperature for several fields (left);
right: as a function of the field for temperatures \{0.05, 0.2, 0.4, 0.6\}.
}
\end{figure}
We start by showing the temperature and field dependence of the
heat conductivity tensor for the $d_{x^2-y^2}$ gap. The heat
current and the field are along the $\vx$-axis, along the gap
maximum, as schematically shown in Fig.~\ref{fig:PDcylK}. The
longitudinal thermal conductivity is seen in Fig.~\ref{fig:Kbt} to
rapidly decrease below $T_c(H)$ (left panel), as the gap opens in
the single-particle spectrum. Notice, however, that the lines
plotted for different fields intersect, implying that the field
dependence of $\kappa_{xx}$ is non-monotonic, as shown in the
right panel. $\kappa_{xx}(H)$ increases with field at the lowest
$T$ and $H$. In this regime the low energy quasiparticles are
located near the gap nodes, where order parameter vanishes,
$\cY(\phi_n)=0$, and the transport lifetime, Eq.(\ref{eq:tauH}) is
limited only by the impurity scattering, $- \Im \Sigma^R$. Hence
in the competition between the increased number of heat-carrying
quasiparticles due to field and scattering on the vortices, the
density of states wins, and the conductivity increases with field.
In contrast, at higher $T$, the unpaired quasiparticles are
already induced by temperature away from the nodes, and turning on
the field leads to increased scattering, hence the decrease in the
thermal conductivity. The evolution of $\kappa_{xx}$ with $T$ and
$H$ is nearly identical to that found for the field normal to the
layers in a vortex state model with a single Landau level
\cite{IVekhter:1999}, and is in agreement with experimental
results \cite{YMatsuda:2006}.

\begin{figure}[t]
\centerline{\includegraphics[height=4cm]{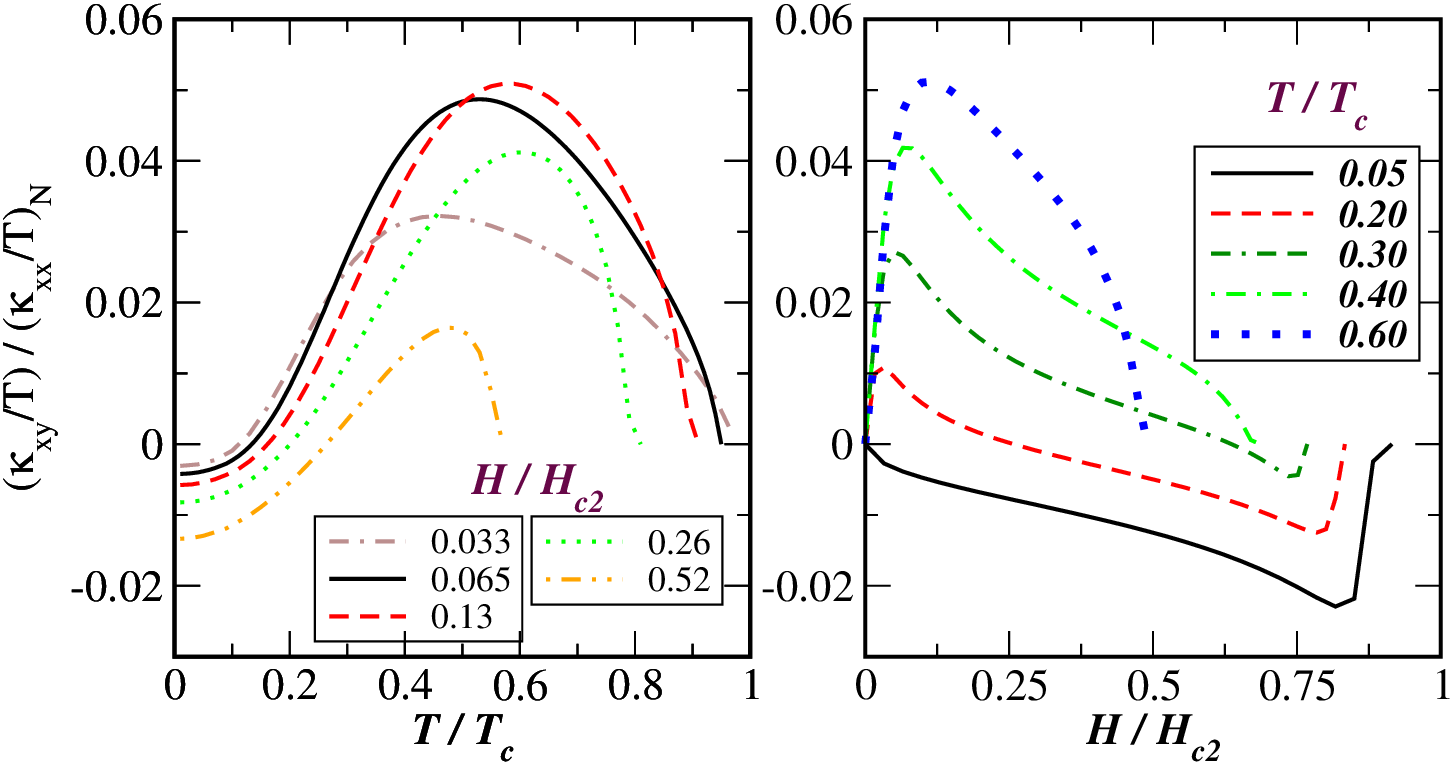}}
\caption{\label{fig:Kxybt} (Color online) Transverse heat conductivity for
$\grad T || \hat{x} $-antinodal direction: $\vH || node$
as a function of temperature (left) and field (right).
}
\end{figure}
Fig.~\ref{fig:Kxybt} shows the temperature (left panel) and field
(right panel) dependence of the transverse heat conductivity. As
we emphasized above, for a general orientation of the field the
dominant contribution to the transverse heat conductivity is not
caused by bending of the quasiparticle trajectories in the applied
field due to Lorenz force, but is due instead to the anisotropic
scattering of quasiparticles by the vortices. If the thermal
gradient is not along one of these two ``transport axes'',  along
the vortices and perpendicular to the vortices, transverse current
arises.  On the other hand, if the thermal gradient is along or
normal to the field direction, $\kappa_{xy}=0$ by symmetry, as is
clear from Eq.~(\ref{eq:symKevenodd}). Hence we show the
transverse conductivity for $\phi_0=45^\circ$ which corresponds to
$\vH || node$.

The transverse conductivity is allowed to change sign, as it only
reflects the difference between the thermal conductivities
parallel and perpendicular to {\bf H}, which themselves depend on
the temperature and field. The temperature dependence shows a
large peak at intermediate $T$ for all $H$ and tends to zero as
the normal state is approached. The field dependence is more
interesting. At low temperature, $\kappa_{xy}$ is negative and
monotonically decreasing up to the critical field, and rapidly
goes to zero at $H_{c2}$. At higher temperature $\kappa_{xy}$ is
positive, and has a peak at low fields.
The temperature, at which the peak first appears, seems correlated
with that where the downturn in $\kappa_{xx}(H)$ is first seen,
and the field value at the peak position moves in step with the
minimum of $\kappa_{xx}(H)$ in Fig.~\ref{fig:Kbt}~(right). It is
therefore likely that this feature is a signature of the increased
scattering due to magnetic field. This is supported by
correlations between the peak and significant
$\kappa(90^\circ)-\kappa(0^\circ)$, which stems from magnetic
scattering.

\begin{figure}[t]
\centerline{\includegraphics[height=5.5cm]{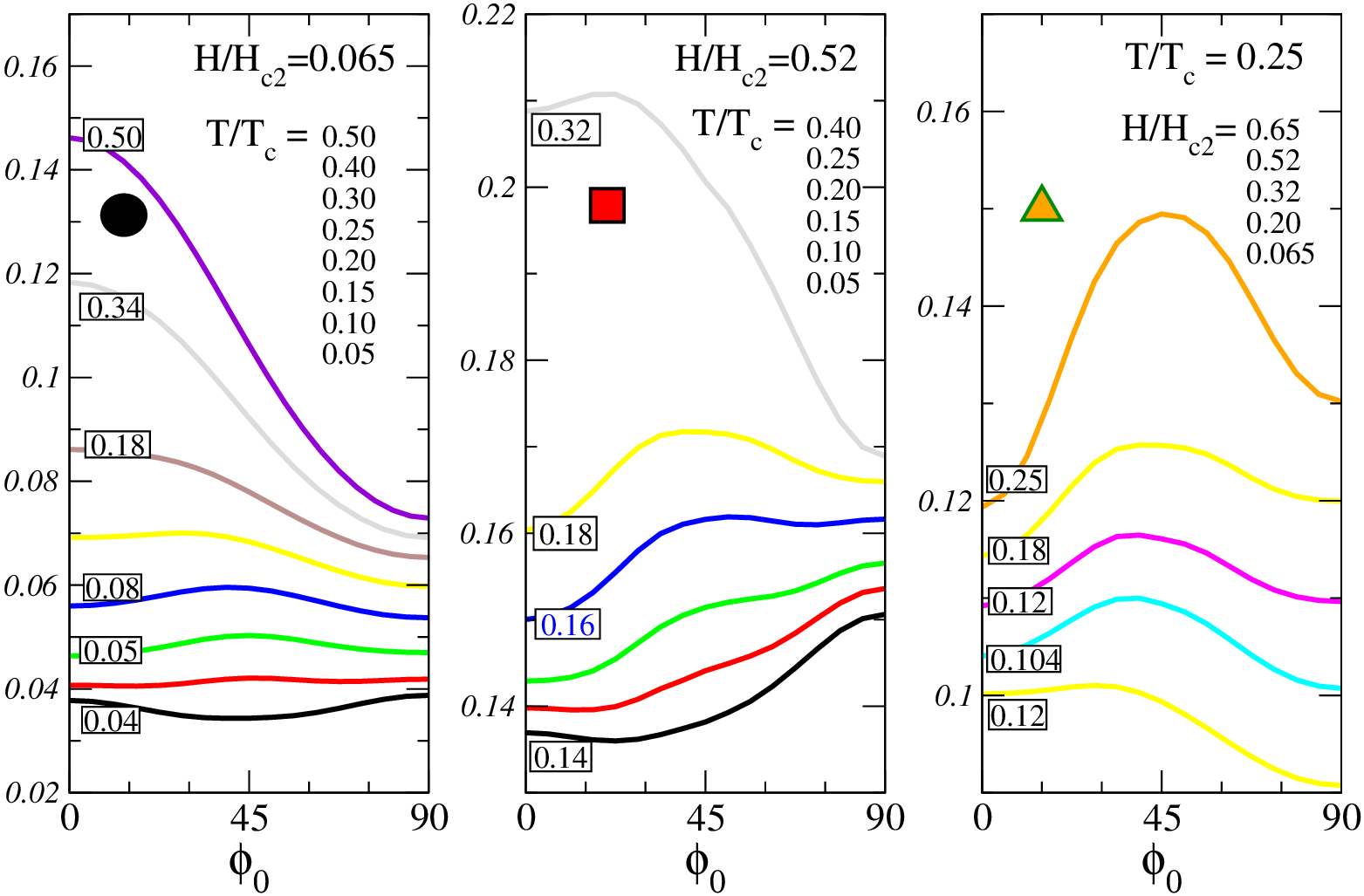}}
\caption{\label{fig:KprofDx2y2} (Color online) Heat conductivity
profiles for $T$-scans at two different fields and a $H$-scan at
$T/T_c=0.25$. Different symbols correspond to points in
Fig.~\ref{fig:PDqcylK}~(left). The curves are shifted vertically
to fit on the same plot, and the value for  $\phi_0=0$ for each of
the shifted curves is given in the corresponding boxes. }
\end{figure}
To make connection to experiment, in Fig.~\ref{fig:KprofDx2y2} we
show the temperature scans of the longitudinal heat conductivity
as a function of the field direction for low and moderate fields
(left and middle panel respectively), and a field scan at
$T/T_c=0.25$ (right panel). The evolution with temperature at low
field is similar to that found for the cylindrical FS, see
Fig.~\ref{fig:PDcylK}. The low temperature region is dominated by
the evolution of the four-fold term, while, as temperature
increases, the two-fold component becomes more prominent. The
field scan  strongly resembles the analogous result for the
cylindrical FS: note the appearance of a pronounced peak for the
nodal direction ($\phi_0=45^\circ$) with increasing field. This
shape of $\kappa_{xx}(\phi_0)$  strongly resembles the
experimentally found anisotropy in CeCoIn$_5$ as shown in
Ref.\onlinecite{KIzawa:CeCoIn5}. This speaks in favor of the
$d_{x^2-y^2}$ gap symmetry in this material, and we will provide
the detailed analysis at the end of this section.

\begin{figure}[t]
\centerline{\includegraphics[height=4.0cm]{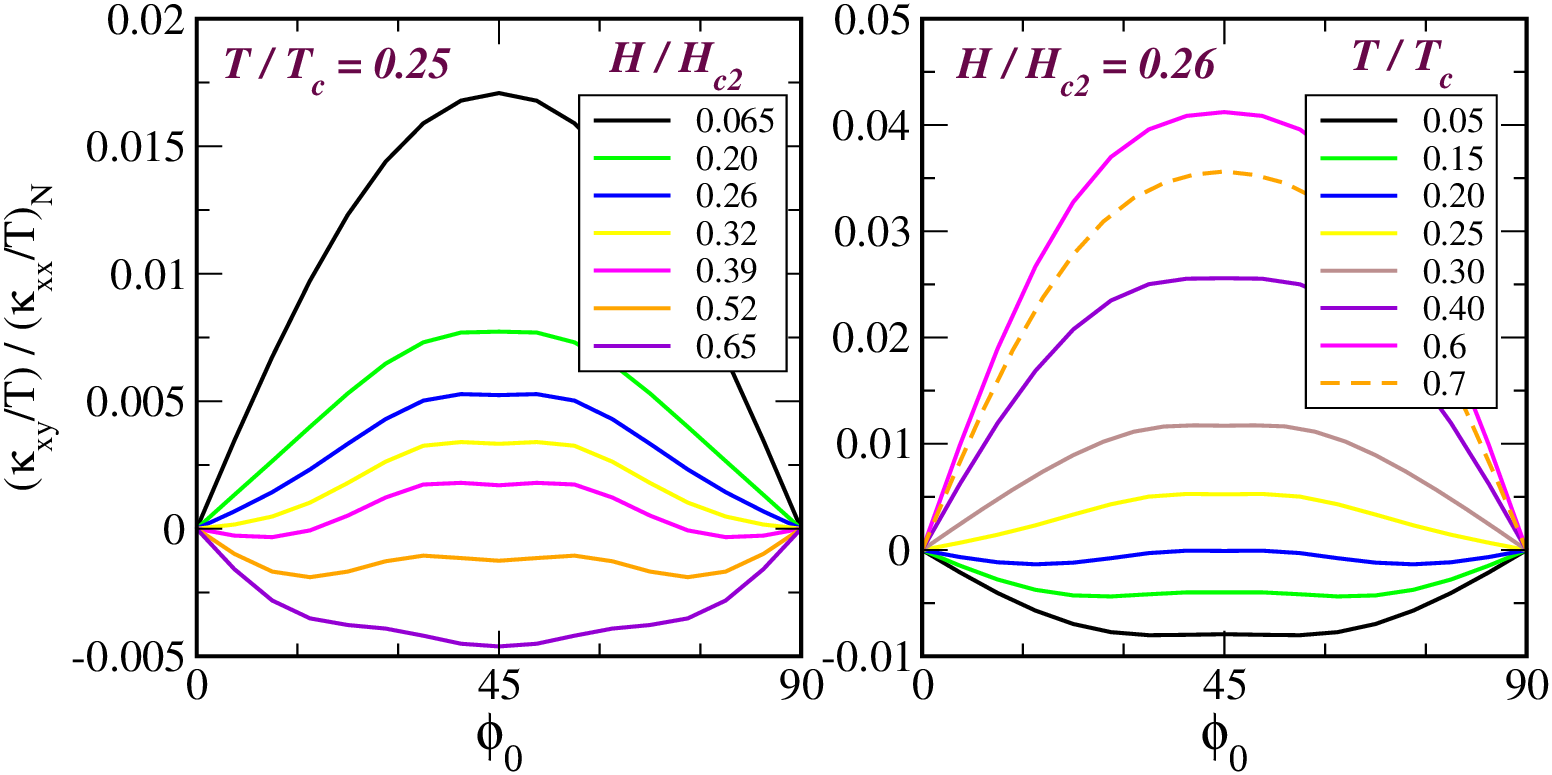}}
\caption{\label{fig:Kxy} (Color online) Anisotropy profile of the
transverse heat conductivity. This component gives the heat
current along the $y$-direction, when the $\grad T || x$ and the
field makes angle $\phi_0$ with the antinodal $x$-axis. }
\end{figure}
For the same relative orientation of the gap and the heat current,
we show a typical profile of the transverse thermal conductivity,
$\kappa_{xy}$, in Fig.~\ref{fig:Kxy}. For $\phi_0=0^\circ$ and
$\phi_0=90^\circ$ this component vanishes identically, see
discussion above. Over a wide range of $T$ and $H$ parameter range
this component shows essentially $\sin2\phi_0$-like behavior
Eq.(\ref{eq:Fourier_x2y2}) that agrees with experimental findings
in high-$T_c$ materials.\cite{ROcana:2001,ROcana:2002} We
emphasize that this modulation is completely unrelated to the
nodal structure of the gap, moreover, only the deviation from the
pure sinusoidal profile, seen in several curves in
Fig.~\ref{fig:Kxy}, carries information about the gap structure.

\begin{figure}[t]
\centerline{\includegraphics[height=5.5cm]{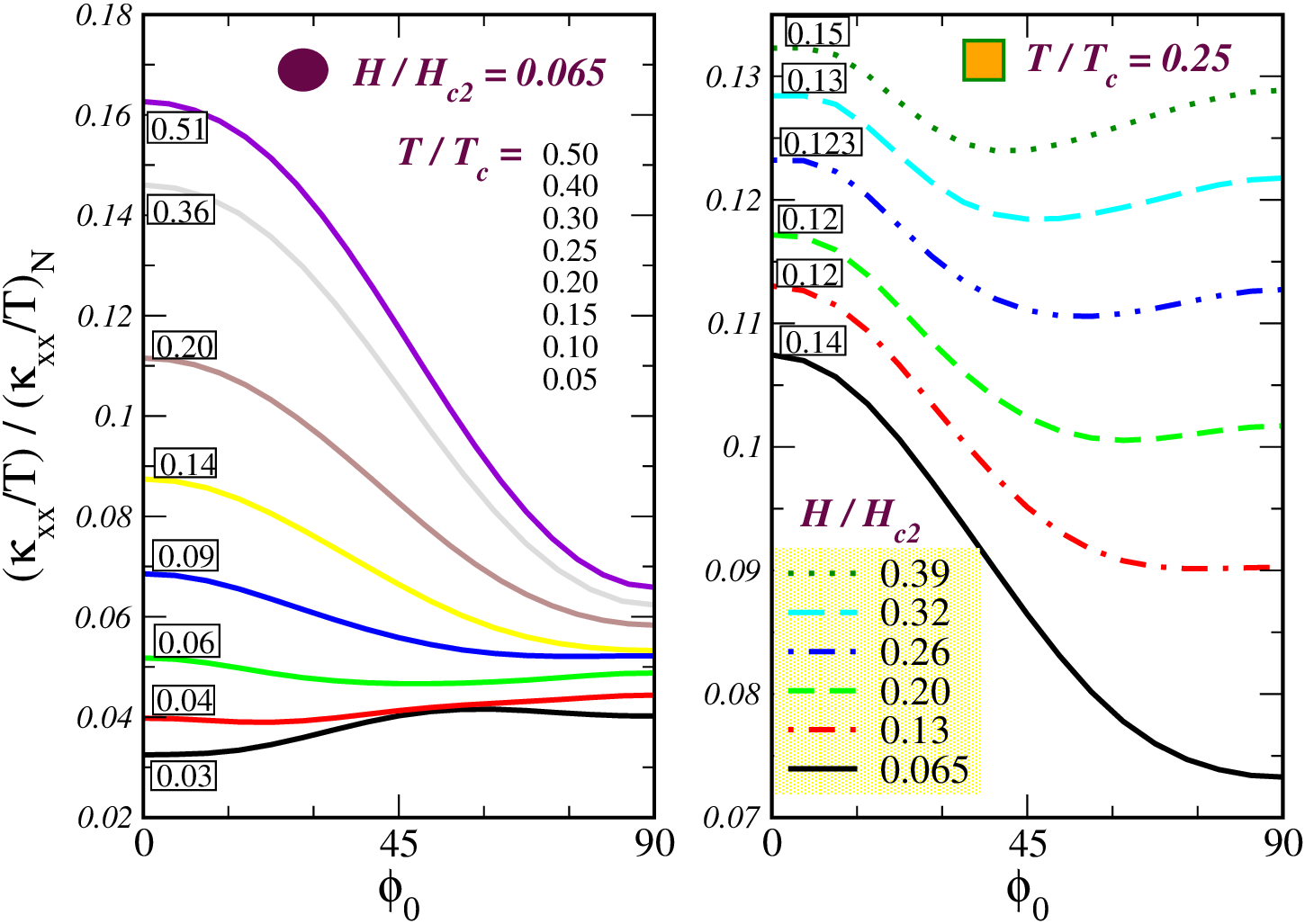}}
\caption{\label{fig:KprofDxy} (Color online)
Longitudinal heat conductivity anisotropy for $d_{xy}$ symmetry.
The curves are shifted vertically, so the scale on the vertical
axis shows only the absolute anisotropy. The relative anisotropy
can be deduced from the true values of $\kappa_{xx}(0)$ shown in boxes.
On the left is shown a temperature scan for $H/H_{c2} = 0.065$,
and on the right is a field scan for $T/T_{c} = 0.25$,
shown correspondingly by filled circles and squares
in Fig.~\ref{fig:PDqcylK}~(right).
}
\end{figure}
To complete the description of the thermal conductivity, we
present in Fig.~\ref{fig:KprofDxy} the longitudinal thermal
conductivity $\kappa_{xx}(\phi_0)$ for $d_{xy}$ symmetry of the
order parameter (when the temperature gradient is along a nodal
direction), and compare it with the results for the $d_{x^2-y^2}$
gap in Fig.~\ref{fig:KprofDx2y2}. The temperature scans for field
$H/H_{c2}=0.065$ in Fig.~\ref{fig:KprofDx2y2}(a) and
Fig.~\ref{fig:KprofDxy}(left) demonstrate that at high
temperatures the same two-fold symmetry holds for both gap
symmetries. At low $T$ the fourfold feature disappears slightly
faster for the $d_{xy}$ gap, and, at the lowest $T$, clearly has
the opposite sign for the two gaps; this is in agreement with
Eq.(\ref{eq:Fourier_x2y2}) and Eq.(\ref{eq:Fourier_xy}).
Comparison of the field scans for $T/T_c=0.25$,
Fig.~\ref{fig:KprofDx2y2}(c) and Fig.~\ref{fig:KprofDxy}(right),
shows a rather dramatic difference between the profile of the
thermal conductivity under the rotated field for the two cases. A
local maximum for the field at $\phi_0=45^\circ$ to the heat
current is clearly resolved for $d_{x^2-y^2}$ gap. For the same
field direction either a minimum or no feature is seen for the
$d_{xy}$ symmetry. Recall that at this temperature, $T\ll T_c$,
the thermally excited quasiparticles are still located near the
nodes of the gap, at $\phi_n$. Their scattering on the vortices
depends on the component of the Fermi velocity normal to the
field, and therefore on the sine of the angle between the nodal
and the field directions. Hence the strongest variation in the
scattering occurs as the field sweeps through the nodal direction,
when $|\phi_n-\phi_0|\ll 1$. For the $d_{x^2-y^2}$ gap, with nodes
at $\phi_n=45^\circ+n 90^\circ$ to the direction of the heat
current, this leads to a noticeable feature in the profile of the
thermal conductivity for $\phi_0\simeq\phi_n$. For the $d_{xy}$
order parameter, with $\phi_n=n 90^\circ$,  the rotated field
sweeps through the nodes at the same time as it is either parallel
or normal to the heat current, $\phi_0\simeq 0^\circ, 90^\circ$.
In that case the twofold transport anisotropy due to vortices
masks the nodal signatures, and the signal is largely twofold.
Only in restricted regions of the phase diagram, when the dominant
twofold anisotropy is nearly absent (intermediate fields in the
right panel of Fig.~\ref{fig:KprofDxy}), does the existence of the
nodes affect the profile of $\kappa_{xx}$. This difference between
the behavior of $\kappa_{xx}(\phi_0)$ for the two types of gap
strongly suggests to us that the experimental results for
CeCoIn$_5$ effectively rule out the $d_{xy}$ symmetry for this
compound.

\begin{figure}[t]
\centerline{\includegraphics[height=5.5cm]{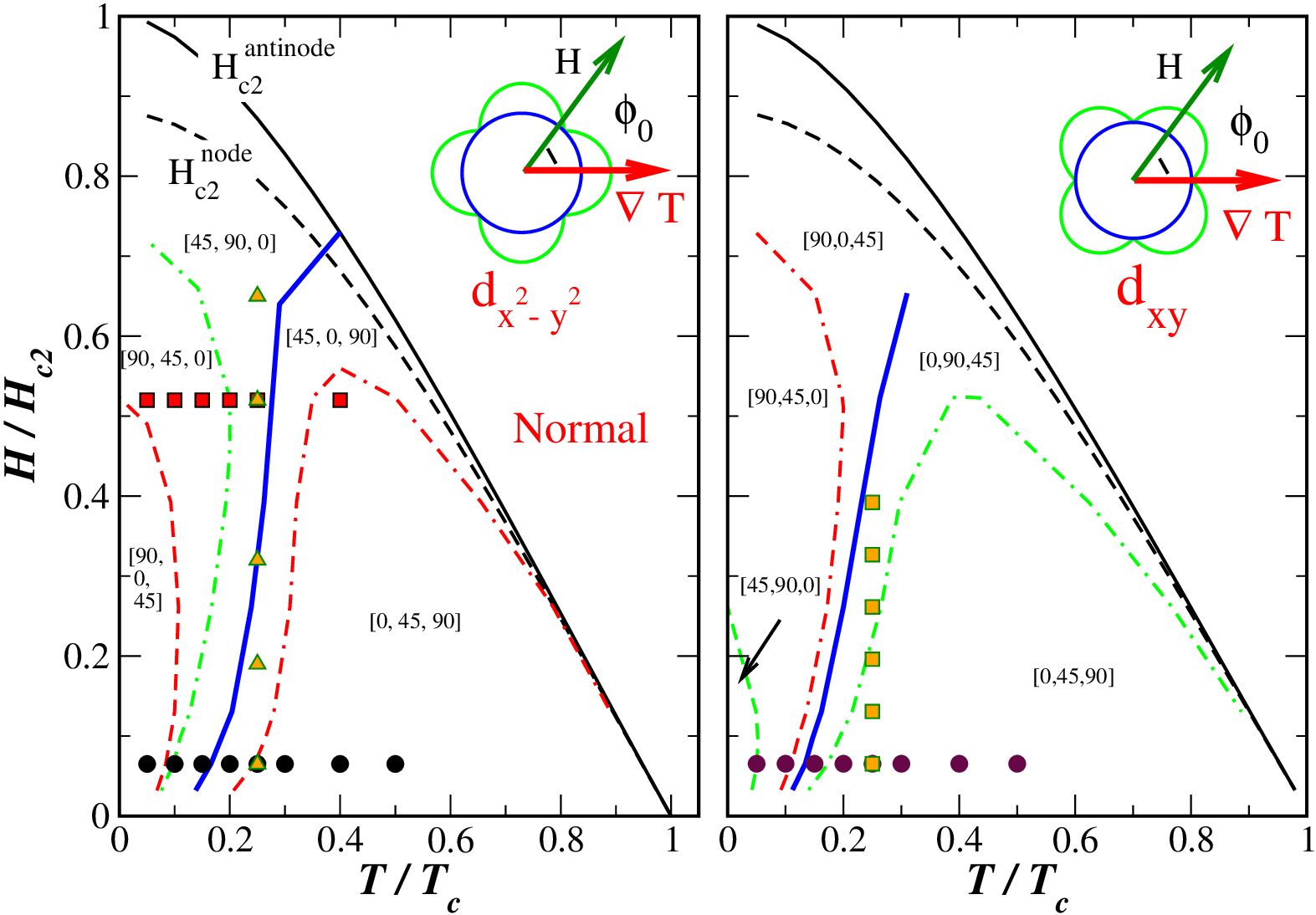}}
\caption{\label{fig:PDqcylK} (Color online) The anisotropy
phasediagram of the longitudinal heat conductivity
$\kappa_{xx}(\phi_0)$. Notation $[\phi_0^a,\phi_0^b,\phi_0^c]$ is
short for $\kappa_{xx}(\phi_0^a)
> \kappa_{xx}(\phi_0^b) > \kappa_{xx}(\phi_0^c)$. The solid blue
line corresponds to the change of sign in the $\kappa_{2}$, which
is positive at high $T$ and negative at low $T$. Its position on
the phase diagram is almost the same for the heat current along
antinode and node. Left panel: the temperature gradient is along
the antinodal direction. The symbols (circles, squares and
triangles) correspond to anisotropy curves in
Fig.~\ref{fig:KprofDx2y2}. The variation of the transverse heat
conductivity anisotropy for this gap orientation is shown in
Fig.~\ref{fig:Kxy}. Right panel: the temperature gradient is along
a nodal direction. }
\end{figure}

In Fig.~\ref{fig:PDqcylK} we summarize the results in the  form of
a phase diagram. The most noticeable differences with the
cylindrical Fermi surface, Fig.~\ref{fig:PDcylK}, occur near the
upper critical field due to the $H_{c2}$ anisotropy, absent in the
non-self-consistent calculation. Away from the critical field, in
the low-to-moderate $T$,$H$ corner, however, the anisotropy shows
very similar features for the cylindrical and the corrugated FS,
although the detailed positions of  the separation lines,
indicating the change in the shape of the thermal conductivity
profile, is different. We believe that the location of these lines
is determined by the symmetry and the shape of the Fermi surface,
and other microscopic details of the material. Recall that the
coupling between different Landau level components of the order
parameter is generated by the action of the differential operator,
${\bf v}_f(\vp)\cdot{\bm \nabla}$, which explicitly depends on the
symmetries of the Fermi velocity. On the other hand, based on the
similarities between the phase diagram computed with the lowest
Landau level, Fig.~\ref{fig:PDcylK}, and that for three
components, Fig.~\ref{fig:PDqcylK}, we conclude that the salient
features and changes in the anisotropy as a function of
temperature and field are captured here.

\begin{figure}[t]
\centerline{\includegraphics[height=3.5cm]{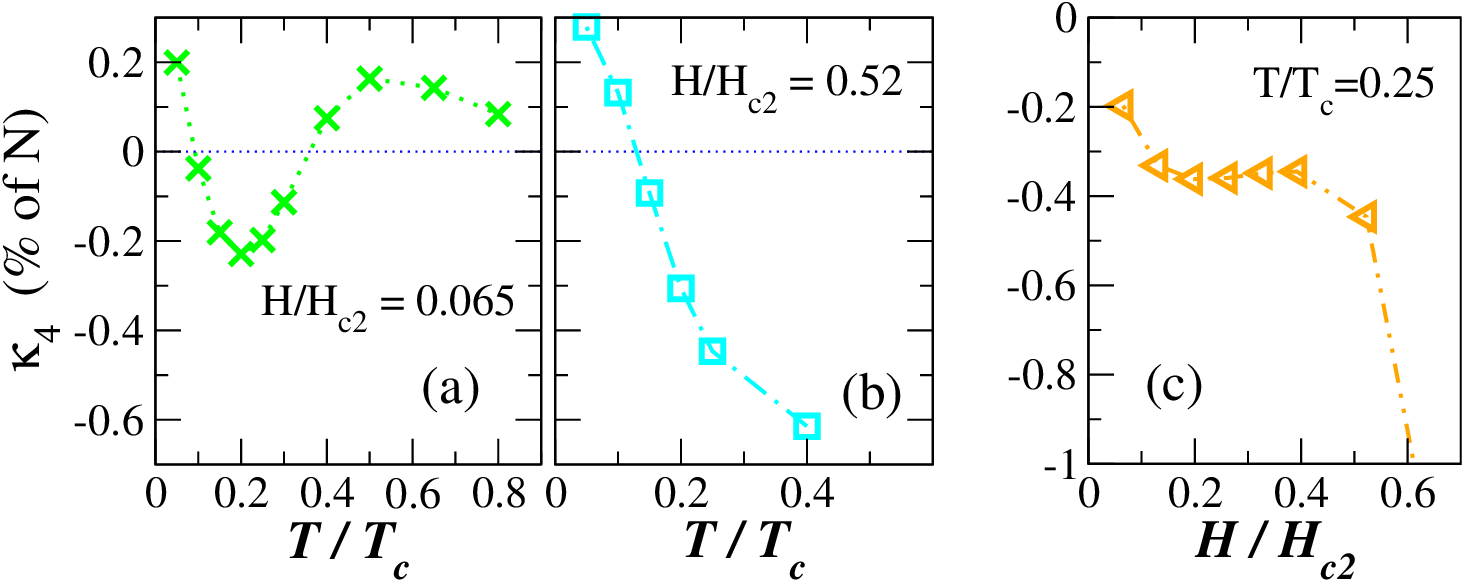}}
\caption{\label{fig:K4} (Color online) The four-fold symmetry coefficient $\kappa_{4}$
of heat conductivity for the scans shown above in figure \ref{fig:KprofDx2y2}.
}
\end{figure}

\begin{figure}[t]
\centerline{\includegraphics[height=5.0cm]{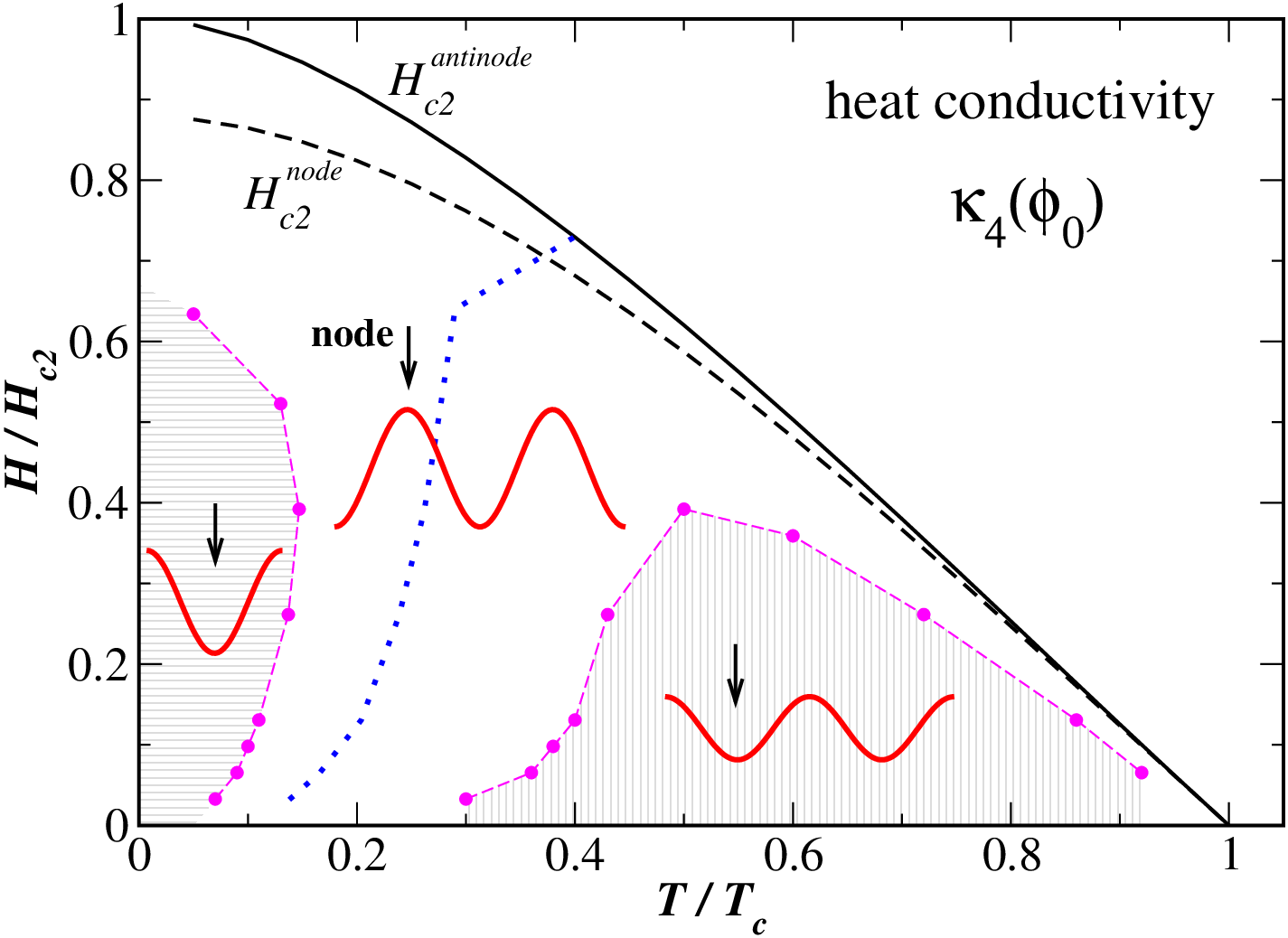}}
\caption{\label{fig:PDK} (Color online) The $T$-$H$ phase diagram
for the fourfold component, $\kappa_4$, of the thermal
conductivity under rotated magnetic field. Differently shaded
regions correspond to different sign of the four-fold term, and
its connection to the nodal directions. The dashed line marks the
crossover from $\kappa_{xx}(0) < \kappa_{xx}(90^\circ)$ at low $T$
to $\kappa_{xx}(0) > \kappa_{xx}(90^\circ)$ at high $T$. }
\end{figure}

It is also clear from the phase diagrams in Fig.~\ref{fig:PDqcylK}
and the anisotropy profiles in Figs.~\ref{fig:KprofDx2y2} and
\ref{fig:KprofDxy} that there is no simple relation between
general shape and evolution of $\kappa_{ij}(\phi_0)$ for the two
symmetries of the gap, $d_{x^2-y^2}$ and $d_{xy}$. On the other
hand, as we noted earlier, the coefficient $\kappa_4$ in the
Fourier decomposition analogous to Eq.~(\ref{eq:Fourier_x2y2}) and
suggested in Ref.~\onlinecite{YMatsuda:2006}, $\kappa_{xx}(\phi_0)
= \kappa_0 + \kappa_2 \cos2\phi_0  + \kappa_4 \cos4\phi_0$,
depends only on the orientation of the field with respect to the
nodes. In Fig.~\ref{fig:K4} we plot the four-fold coefficient
$\kappa_4$ for the values of $T$ and $H$ shown in
Fig.~\ref{fig:KprofDx2y2} for the $d_{x^2-y^2}$ gap.
The four-fold anisotropy changes sign at intermediate temperatures
for both small and moderate fields, see left two panels of
Fig.~\ref{fig:K4}. The coefficient $\kappa_4$ is small near $T_c$,
where only the two-fold pattern that reflects the difference
between transport parallel and normal to the applied field is
detectable. The second sign change in the low-field range, left
panel of Fig.~\ref{fig:K4}, occurs at low temperature, close to
the limit of validity of the BPT approximation
\cite{AVorontsov:I}. However, it is this feature that is connected
in the phase diagram to the reversal shown in the middle panel for
higher fields, see Fig.~\ref{fig:PDK}, which suggests that it is
not an artifact of the approach, but a real effect.
For the field scan at $T/T_c=0.25$ (right panels of
Figs.~\ref{fig:KprofDx2y2},\ref{fig:K4}) $\kappa_4$ is always
negative, so that the minima in the four-fold component mark the
antinodal directions, while the maxima occur for $\vH$ along the
nodes. There is a sharp increase in the magnitude of this
coefficient as we approach the critical field due to the in-plane
$H_{c2}$ anisotropy.

\subsection{Comparison with experiment: CeCoIn$_5$ and
$\kappa$-(BEDT-TTF)$_2$Cu(NCS)$_2$}

We close this section with a detailed comparison of our results
with the experimental data.  One of the main motivations for this
work was the measurement of the thermal conductivity in a heavy
fermion  CeCoIn$_5$, ~\cite{KIzawa:CeCoIn5}. Another example of
quasi-2D superconductor where the anisotropy was measured is
$\kappa$-(BEDT-TTF)$_2$Cu(NCS)$_2$.~\cite{KIzawa:BEDT}.

In CeCoIn$_5$ the heat current is driven along [100] crystal
direction. The observed profile of the heat conductivity at
$T/T_c=0.25$ is in good agreement with that shown in
Fig.~\ref{fig:KprofDx2y2} ($d_{x^2-y^2}$) for comparable
temperature ($T \approx 0.2 T_c$), including the peak for the
field at 45$^\circ$ to the heat current. The profile differs
significantly from that expected for a $d_{xy}$ gap as shown in
Fig.~\ref{fig:KprofDxy}. We find that the behavior of the
experimentally determined four-fold term amplitude, $\kappa_4$,
agrees with Fig.~\ref{fig:K4}~(right): it vanishes as $H \to 0$
and saturates at $H \geq 0.2 \, H_{c2}$. The overall amplitude of
this component is smaller  in our computation than that observed
experimentally by approximately a factor of three: however, since
this magnitude is determined by the shape of the Fermi surface, we
do not expect the model calculation to be quantitatively correct.
Our results at moderate temperatures are consistent with the
experimental temperature scan $T = 0.15 - 0.9 \, T_c$ at $H
\approx 0.1 H_{c2}$. At $T\sim T_c$ our results suggest an
inversion of the $\kappa_4$-term which was not observed. However,
in this region CeCoIn$_5$ still has strong inelastic scattering
(resulting in a peak of the thermal conductivity at $T\sim 0.75
T_c$) which was not included in the calculation. Experimentally,
extraction of the small $\kappa_4$ amplitude on the background of
the dominant twofold term has greater relative errors. Moreover,
since in this range $\kappa_0\gg\kappa_2\gg\kappa_4$, the
difference between the calculated and the measured heat
conductivity described in Sec.~\ref{sec:calc_vs_measur} may also
contribute to the discrepancy. Finally, since the upper critical
field in CeCoIn$_5$ is paramagnetically limited, we cannot make a
reliable connection of our results with experiment at low
temperatures and high fields; in contrast, Zeeman splitting does
not affect the low-to-intermediate $T$-$H$ behavior. Therefore
reliable comparison can be made only in the region away from the
critical field, where a maximum in the fourfold component points
to the node, strongly implying $d_{x^2-y^2}$ symmetry in agreement
with Ref.~\onlinecite{KIzawa:CeCoIn5}.  Note that, according to
our analysis, generally the line of inversion of the fourfold term
in Fig.~\ref{fig:PDK} is distinct from the line separating the
increasing and decreasing $\kappa(H,T)$ at low fields, which was
used to decide whether the minima or the maxima of the
oscillations indicate the nodes in
Ref.~\onlinecite{KIzawa:CeCoIn5}.

In the quasi-2D organic superconductor
$\kappa$-(BEDT-TTF)$_2$Cu(NCS)$_2$ the available data are in
low-$T,H$ region only~\cite{KIzawa:BEDT}.  The heat current is
driven along [110] axis. Extensive analysis of the experimental
data is required to separate the electronic contribution (which is
small due low carrier density), from the phonon heat transport.
If, however, we concentrate our attention only on the behavior of
the four-fold electronic term, the observed anisotropy fits well
into the phase diagram for the $d_{xy}$ gap,
Fig.~\ref{fig:PDqcylK}. In the region $H \sim 0.07 H_{c2}$, $T =
0.04 - 0.07 T_c$, the fourfold term, $\kappa_4 \cos 4\phi_0$, has
a maximum for  $\phi_0 = 0^\circ$ (along [110]) for $T\sim 0.04
T_c$, and essentially disappears at $T\sim 0.07 T_c$. In our
mapping to the phase diagram, in the experimental regime the field
along the node produces minima in the conductivity, and we concur
with Ref.~\onlinecite{KIzawa:BEDT} that the nodal direction is
[100], which suggests the $d_{xy}$ symmetry.

\section{\label{sec:CON} Conclusions}

In this paper, following our approach in I for the specific heat,
and using the non-equilibrium Keldysh formulation of the
quasiclassical theory,  we derived a general expression for the
heat conductivity tensor of a superconductor in magnetic field.
The derivation was based on the closed form solution for the
Green's function obtained in I, that made use of the
Brandt-Pesch-Tewordt approximation. The utility of this approach
lies in its ability to self-consistently take into account
impurity scattering, the detailed shape of the Fermi surface, and
multiple Landau levels in the order parameter in the vortex state.
Numerical computations based on this approach are very
time-efficient. The main advantage of our approach is that it
provides unified method for calculations of transport and
thermodynamics over a large range of temperatures and fields, well
beyond the realm of applicability of the  semiclassical schemes.

\begin{figure}[t]
\centerline{\includegraphics[height=5.0cm]{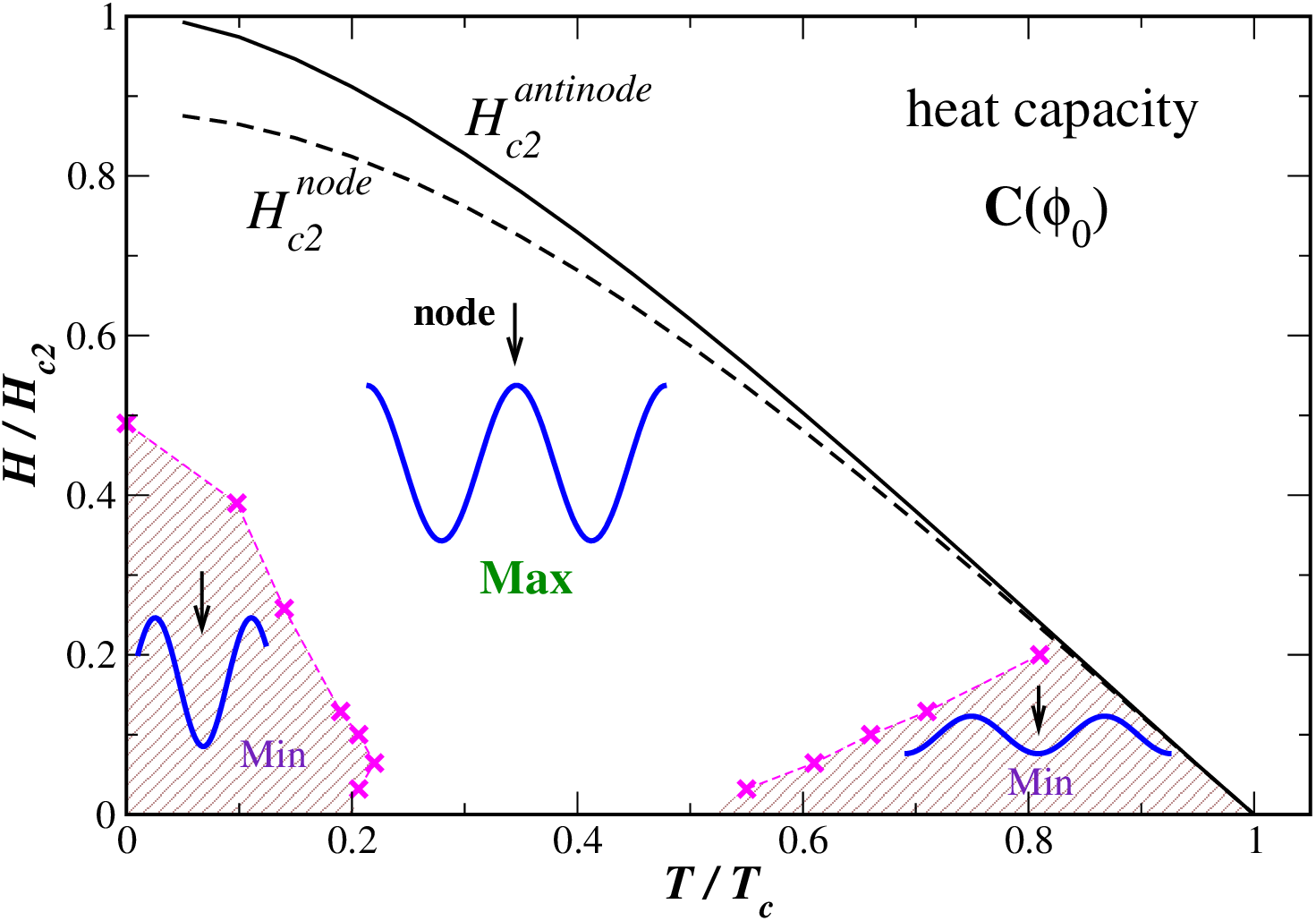}}
\caption{\label{fig:PDC} (Color online)
The $T$-$H$ phase diagram for anisotropy of the heat capacity
under rotated magnetic field. We choose for our model a
quasi-cylindrical Fermi surface and a $d$-wave order parameter.
For rotations of the field in the $ab$-plane the oscillations
of the heat conductivity are four-fold, and a different feature
(maximum or minimum) of this dependence determine the
node position (arrows), depending on the location on this diagram.
}
\end{figure}

In these two papers we applied the developed formalism to a
$d$-wave superconductor with a quasi-cylindrical Fermi surface. We
concentrated on the behavior of the specific heat and thermal
conductivity, since they are the most widely used experimental
probes. To make a connection between theory and experiment, we
provided, for the first time, a complete description of the
anisotropy of the thermal conductivity and specific heat in the
$T$-$H$ phase diagram, starting from the ``semiclassical'' region
at low $T,H$ and up to the critical field.

Two figures summarize the main results of this work. Figure
\ref{fig:PDC} recalls the results of I, and shows the phase
diagram for anisotropy of the specific heat under rotated magnetic
field, $C(\phi_0)$. The preceding figure, Fig.~\ref{fig:PDK},
shows the anisotropy of the fourfold component, $\kappa_4$, of the
longitudinal thermal conductivity, $\kappa_{xx}$, for the model.
One of our main findings is that both anisotropic signatures
change sign, i.e. invert, in the $T$-$H$ plane. For the specific
heat, the inversion of the anisotropy is due to the effect of the
quasiparticle scattering on vortices on the density of states, see
I. The semiclassical (Doppler shift) picture predicting a minimum
of $C$ for the field along a node is valid at low $T$ and $H$,
where it was designed to work. The effects of the energy shift,
however, are superseded by the redistribution of the spectral
weight due to scattering, which becomes dominant not only at
moderate fields, but also at low fields for finite energies.
Consequently, the anisotropy changes sign at finite $T$,
Fig.~\ref{fig:PDC}.

Analysis of the heat conductivity is more involved due to
interdependence of the transport scattering time and the density
of states in the self-consistent treatment. We showed that, under
a rotated field, the four-fold term in Fourier decomposition of
the heat conductivity, $\kappa_4(\phi_0)$, exhibits signatures of
the nodes, and depends only on the angle between the field, $\vH$,
and the nodal directions, but not on the orientation of the heat
current relative to the nodes. From comparison of
Fig.~\ref{fig:PDK} and Fig.~\ref{fig:PDC} it is clear that the
evolution of the fourfold coefficients in the specific heat and
thermal conductivity, including sign changes, is quite similar
across the phase diagram.

The exact location of the inversion lines depends on the
microscopic details, such as the Fermi surface shape and the
detailed structure of the order parameter. The relative position
of these lines, however, is stable with respect to the moderate
changes of the FS curvature along $z$-axis and impurity
concentration.  The developed theory is valid over most of the
phase diagram, except for very low fields in dirty samples, where
the averaging procedure in the Brandt-Pesch-Tewordt approximation
is no longer valid.  Thus the qualitative changes in the fourfold
term at moderate fields, where the anisotropy is the most
prominent, should be detectable experimentally (albeit may prove
labor-intensive).  One possible experimental approach to search
for the node locations would be to measure anisotropy at several
points in the phase diagram, in order to map out the evolution of
the anisotropic contribution.

Finally, by comparing our results with available experimental data
on the specific heat and thermal conductivity we concluded that
the order parameter of heavy fermion material
CeCoIn$_5$\cite{KIzawa:CeCoIn5,HAoki:2004} has $d_{x^2-y^2}$
symmetry, and reconciled the thermodynamic and transport
measurements. Analysis of the thermal conductivity for organic
superconductor
$\kappa$-(BEDT-TTF)$_2$Cu(NCS)$_2$\cite{KIzawa:BEDT} places it
most likely into $d_{xy}$ family. We believe that our method,
which allows detailed microscopic calculations for specific
compounds, will enable unambiguous interpretation of the
anisotropy in thermal and transport properties of unconventional
superconductors, and will lead to maturing of this method as a
tool for determining the nodal directions.

\section{Acknowledgements}

This work was partly done at KITP with support from NSF Grant
PHY99-07949;  and was also supported by the Board of Regents of
Louisiana. We thank D.~A.~Browne, C.~Capan, P.~J.~Hirschfeld,
Y.~Matsuda, and T.~Sakakibara for discussions. I.~V. is grateful
to A. Houghton for encouragement during early stages of this work.

\appendix

\section{\label{app:Keldysh} Details of the heat conductivity derivation}

Our approach allows to obtain an expression for thermal
conductivity that generalizes previous results to the case of
vortex state with multi-Landau level order parameter, and
arbitrary impurity strength.

Keldysh diagram technique is formulated for 8$\times$8 Green's
function,\cite{ser83} which is traditionally split into three
$4\times4$ parts, Retarded (R), Advanced (A) and Keldysh
(K):$\whg^X$, $X=\{R,A,K\}$,
    \be
    \check{g} =
    \left( \begin{array}{cc}
    \whg^R & \whg^K \\ 0 & \whg^A
    \end{array} \right) \,,
    \ee
    \be
    \whg^X(\hat{\vp}, \vR; \vare) =
    \left( \begin{array}{cc}
    g^X + \vg^X \vsigma & (f^X + \vf^X\vsigma) i\sigma_2 \\
    i\sigma_2 (\ul{f}^X + \ul{\vf}^X\vsigma) & \ul{g}^X + \ul{\vg}^X \vsigma^*
    \end{array} \right)
    \ee
For stationary problems these functions obey the normalization
conditions:
    \be
    \whg^{R,A} \whg^{R,A} = -\pi^2 \hat{1} \qquad\,, \qquad
    \whg^{R} \whg^K + \whg^K \whg^A = 0 .
    \ee
We do not need to solve equations for all the functions as they
are related through symmetries,\cite{ser83}
    \be
    \begin{array}{c@{\quad}c}
    g^R(\hat{\vp}; \vare) = g^A(\hat{\vp}; \vare)^* &
    \vg^R(\hat{\vp}; \vare) = \vg^A(\hat{\vp}; \vare)^* \\
    g^R(\hat{\vp}; \vare) = \ul{g}^A(-\hat{\vp}; -\vare) &
    \vg^R(\hat{\vp}; \vare) = \ul{\vg}^A(-\hat{\vp}; -\vare) \\
    g^K(\hat{\vp}; \vare) = \ul{g}^K(-\hat{\vp}; -\vare)  &
    \vg^K(\hat{\vp}; \vare) = \ul{\vg}^K(-\hat{\vp}; -\vare)
    \end{array}
    \ee
    \be
    \begin{array}{c@{\quad}c}
    f^R(\hat{\vp}; \vare) = \ul{f}^A(\hat{\vp}; \vare)^* &
    \vf^R(\hat{\vp}; \vare) = \ul{\vf}^A(\hat{\vp}; \vare)^* \\
    f^R(\hat{\vp}; \vare) = {f}^A(-\hat{\vp}; -\vare) &
    \vf^R(\hat{\vp}; \vare) = {\vf}^A(-\hat{\vp}; -\vare) \\
    f^K(\hat{\vp}; \vare) = -\ul{f}^K(\hat{\vp}; \vare)^*  &
    \vf^K(\hat{\vp}; \vare) = -\ul{\vf}^K(\hat{\vp}; \vare)^*
    \end{array}
    \ee
The different functions obey transport equations:
    \bea
    &&[(\vare - \sigma_B) \widehat{\tau}_3 - \whs^{R,A}, \whg^{R,A}] + i\vv_f \grad \whg^{R,A}=0
    \\
    &&((\vare - \sigma_B) \widehat{\tau}_3 - \whs^{R}) \whg^K -
    \whg^K ((\vare - \sigma_B) \widehat{\tau}_3 - \whs^{A})
    \nonumber \\
    && \hspace{1cm} -\whs^K \whg^R + \whg^A \whs^K + i\vv_f \grad \whg^K = 0 \,.
    \eea
Here
    \be
    \sigma_B =  -{e\over c} \vv_f \vA  \,,
    \ee
is the coupling of quasiparticles to an external magnetic field.
The self-energy is decomposed into the mean-filed order parameter
and the impurity contributions, $\whs^X = \whDelta^X +
\whs^X_{imp}$. Self-consistency equations for the singlet order
parameter are
    \bea
    &&\Delta^R(\vp, \vR) = \Delta^A(\vp, \vR) =
    \\ && \hspace{1cm}
    =\int\limits_{-\infty}^{+\infty} {d\vare\over 4\pi i}
        \int d \hvp' n_f(\hvp') V_s(\vp,\vp') f^K(\vp', \vR;
        \vare)\,,
    \nonumber \\
    &&\Delta^K(\vp, \vR) = 0\,,
    \eea
and for triplet superconductivity the equations are identical, upon
replacing $\Delta^X$ by its vector counterpart, $\vDelta^X$.
The Keldysh part of the self-energy comes in this case from impurities
only. The self-consistent t-matrix approximation for isotropic
impurity scattering gives
    \bea
    &&\whs^X_{imp}(\vR; \vare) = n_{imp} \widehat{t}^X(\vR; \vare)
    \\
    &&\widehat{t}^{R,A} = u \hat{1} + u N_f \langle \whg^{R,A} \rangle \widehat{t}^{R,A}
    \label{eq:TmatrRA}
    \\
    &&\widehat{t}^K = N_f \widehat{t}^{R} \langle \whg^{K} \rangle \widehat{t}^{A}
    \eea
where the angular brackets denote the normalized Fermi surface
average as in the main text of the paper.

In this appendix we denote functions in thermal equilibrium by the
subscript `eq', but in the main text we omit it for brevity. In
local equilibrium,
    \bea
    \whg_\sm{eq}^K = \whg_\sm{eq}^R \Phi_\sm{eq} - \Phi_\sm{eq} \whg_\sm{eq}^A \; , \;
    \whs_\sm{eq}^K = \whs_\sm{eq}^R \Phi_\sm{eq} - \Phi_\sm{eq} \whs_\sm{eq}^A \,,
    \\
    \Phi_\sm{eq} = \tanh \left( {\vare\over 2 T(\vR)} \right) \,.
    \eea
Heat current is
    \be
    \vj_h(\vR) = 2 N_f \int d\vp_f n_f(\vp_f) \int\limits_{-\infty}^{+\infty} {d\vare\over 4\pi i }
    \vare \vv_f(\vp_f) g^K(\vp_f, \vR; \vare) \,,
    \ee
where $g^K(\vp_f, \vR; \vare)$ is the diagonal part of $\whg^K$.
Using the Green's function symmetries one can show that in this
formula it is equivalent to $g^K =  1/4\, \mbox{Tr}(\whg^K)$. The
factor of two reflects our definition of $N_f$ for single spin
projection. Thermal conductivity in the linear response is
determined from
    \bea
    &&\delta \vj_h(\vR) \equiv -\hat{\kappa} \, \grad T =
    \\
    && = 2 N_f \int d\vp_f n_f(\vp_f) \int\limits_{-\infty}^{+\infty} {d\vare\over 4\pi i }
    \vare \vv_f(\vp_f) \, {1\over4} \mbox{Tr} [ \delta g^K(\vp_f, \vR; \vare)] \,.
    \nonumber
    \eea

Our goal is to find the linear, in the temperature gradient that
drives the heat current, correction to the Keldysh Green's
function, $\delta g^K$. The Green's functions varies on the scale
of the magnetic length (or intervortex distance), $\Lambda$, and
on the scale of the superconducting coherence length, $\xi_0$. We
assume that $\Lambda, \xi_0 \ll L_T$, where $L_T$ is the length
scale for temperature variation. We now write the gradient term as
a sum of the gradients due to inhomogeneity and due to the
externally imposed slow temperature variation,
    \be
    \grad = \grad_x + \grad T \; \frac{\partial}{\partial T} \,,
    \ee
with the last term much smaller than the first.

Solution in the local equilibrium is obtained from
    \bea
    [\widehat{h}_\sm{eq}^{R,A}, \whg_\sm{eq}^{R,A}] + i\vv_f \grad_x \whg^{R,A}_\sm{eq} = 0
    \nonumber \\
    \widehat{h}_\sm{eq}^{R,A} = (\vare -\sigma_B) \widehat{\tau}_3 - \whs^{R,A}_\sm{eq} \,.
    \eea
We note that the following analysis can be easily adjusted if we
add an external potential to $\widehat{h}_\sm{eq}^{R,A}$. For now
we continue without it and write down linearized equations for
Green's functions near equilibrium, $\whg^X = \whg^X_\sm{eq} +
\delta\whg^X$, with driving term due to $\vv_f \grad_T \equiv
\vv_f \, \grad T \, \partial_T$,
\begin{widetext}
    \bea
    &&[\widehat{h}_\sm{eq}^{R,A}, \delta \whg^{R,A}]+[-\delta \whs^{R,A}, \whg_\sm{eq}^{R,A}] +
    i\vv_f\grad_x (\delta \whg^{R,A}) + i \vv_f\grad_T \, \whg_\sm{eq}^{R,A} = 0
    \\
    &&\widehat{h}_\sm{eq}^R \delta\whg^K -\delta\whg^K \widehat{h}_\sm{eq}^A -\delta\whs^R \whg_\sm{eq}^K + \whg_\sm{eq}^K \delta\whs^A
    -\delta\whs^K \whg_\sm{eq}^A + \whg_\sm{eq}^R \delta\whs^K
    - \whs_\sm{eq}^K \delta\whg^A + \delta\whg^R \whs_\sm{eq}^K
    \nonumber \\
    &&\hspace{8cm} + i\vv_f\grad_x (\delta \whg^K) + i \vv_f\grad_T \, \whg_\sm{eq}^K = 0
    \eea
\end{widetext}
We decouple equations for $\delta\whg^{R,A}$ from $\delta\whg^K$
by introducing Eliashberg anomalous
propagator\cite{rai95,MJGraf:1996} and the self-energy,
    \bea
    \delta\whg^K &=& \delta\whg^R \Phi_\sm{eq}  - \Phi_\sm{eq} \delta\whg^A +
    \delta\whg^a\,,
    \\
    \delta\whs^K &=& \delta\whs^R \Phi_\sm{eq}  - \Phi_\sm{eq} \delta\whs^A +
    \delta\whs^a\,.
    \eea
The heat current is determined by $\delta\whg^a$, which satisfies
    \bea
    \widehat{h}_\sm{eq}^R \delta\whg^a &-& \delta\whg^a \widehat{h}_\sm{eq}^A + i\vv_f \grad_x \, \delta\whg^a =
    \\
    &=& -i\vv_f \grad_T \Phi_\sm{eq} \; (\whg_\sm{eq}^R - \whg_\sm{eq}^A)
    + \delta\whs^a \whg_\sm{eq}^A - \whg_\sm{eq}^R \delta\whs^a \,.
    \nonumber
    \eea
We need to solve this equation together with the self-consistency
equations on $\delta\whs^X$. Normalization requires
    \bea
    \whg^{R,A}_\sm{eq} \, \delta\whg^{R,A} +  \delta\whg^{R,A} \, \whg^{R,A}_\sm{eq} =
    0\,,
    \label{eq:normRA}
    \\
    \whg_\sm{eq}^R \, \delta\whg^a + \delta\whg^a \, \whg_\sm{eq}^A = 0 \,.
    \label{eq:norma}
    \eea
Up to this point all the equations are valid for both singlet and
triplet pairing states. Below we focus on the singlet pairing and
only briefly comment on the differences between the singlet and
the triplet cases. The self-consistency for the retarded and
advanced linear corrections contain the order parameter and the
impurity contributions, $\delta\whs^{R,A} = \delta\whDelta +
\delta\whs^{R,A}_{imp}$, with
    \bea
    \delta\Delta = \int {d\vare\over 4\pi i} \int d \vp' n_f(\vp')
        V(\vp,\vp') \; \delta f^K(\vp', \vR; \vare)
    \\
    \delta\whs_{imp}^{R,A} = n_{imp} N_f \, \widehat{t}^{R,A}_\sm{eq}
    \langle \delta \whg^{R,A} \rangle \widehat{t}^{R,A}_\sm{eq} \,,
    \eea
but the anomalous part is due to impurities only,
    \be
    \delta\whs_{imp}^{a} = n_{imp} N_f \, \widehat{t}^R_\sm{eq}
    \langle \delta \whg^{a} \rangle \widehat{t}^A_\sm{eq}
    \ee
We see that the equations for the anomalous Green's function,
$\delta\whg^a$, and the self-energy, $\delta\whs^a$, are
completely decoupled from those for the retarded and advanced
Green's functions. On the other hand, the equations for
$\delta\whg^{R,A}$ depend on the anomalous $\delta\whg^a$ through
the variation $\delta\Delta$. For simple retarded and advanced
Green's function ($Y=R,A$),
    \be
    \whg^Y_\sm{eq} = \left( \begin{array}{cc}
    g^Y_\sm{eq}  & f^Y_\sm{eq}  i\sigma_2 \\
    i\sigma_2 \ul{f}^Y_\sm{eq}  & -g^Y_\sm{eq}
    \end{array} \right)
    \ee
we obtain the impurity t-matrix in equilibrium,
\begin{widetext}
    \be
    \widehat{t}^Y_\sm{eq} = \left( \begin{array}{cc}
    t_+ + t_- & t_\Delta \, i\sigma_2 \\
    i\sigma_2 \, \ul{t}_\Delta & t_+-t_-
    \end{array} \right)
    =
    {1\over n_{imp}} \frac{\Gamma \sin^2\delta_0}{1-{\sin^2\delta_0\over \pi^2}
    (\langle g_\sm{eq}^Y\rangle^2-\overline{\langle f_\sm{eq}^Y\rangle \langle \ul{f}_\sm{eq}^Y \rangle} + \pi^2 )}
    \left( \begin{array}{cc}
    \mbox{ctg} \delta_0 + \langle g_\sm{eq}^Y\rangle/\pi & (\langle f_\sm{eq}^Y\rangle /\pi)  \, i\sigma_2 \\
    i\sigma_2 \, (\langle \ul{f}_\sm{eq}^Y \rangle/\pi) & \mbox{ctg} \delta_0 - \langle g_\sm{eq}^Y\rangle/\pi
    \end{array} \right)\,.
    \label{eq:explicit_imp}
    \ee
\end{widetext}
Here we solved (\ref{eq:TmatrRA}) using the BPT approximation.
Note that for the triplet case the off-diagonal parts $\langle
f_\sm{eq}\rangle$ vanish due to inversion symmetry, $\hvp \to
-\hvp$.

For the singlet case we assume $\delta\whs^a = 0$, and validate
this assumption at the end of the calculation. We show that the
linear correction, $\delta\whg^a$, a product of an even, under
$\hvp \to -\hvp$ reflection,  function and an odd, in momentum,
term $\vv_f(\hvp) \grad_T$, so that its average over the FS
vanishes.

We only need to solve the equation for the anomalous part of the
Green's function, since neither Retarded nor Advanced part has a
unit-diagonal term. Hence $\mbox{Tr}[ \whg^{R,A}] = 0$, and they
do not contribute to the heat current. Now $\mbox{Tr}[
\delta\whg^{K}] = \mbox{Tr}[ \delta\whg^{a}]$, and we find
    \bea
    \widehat{h}_\sm{eq}^R \delta\whg^a &-& \delta\whg^a \widehat{h}_\sm{eq}^A
    + i\vv_f\grad_x \delta\whg^a =
    \nonumber \\
    &=& - i \vv_f \grad_T \Phi_\sm{eq} \; (\whg_\sm{eq}^R - \whg_\sm{eq}^A)
    \label{eq:lina}
    \eea
where the matrix,
    \bea
    \widehat{h}_\sm{eq}^{R,A} &=& (\vare - \Sigma^{R,A}_\sm{eq} - \sigma_B) \hat{\tau}_3 - D^{R,A}_\sm{eq} \hat{1} -
    \widehat{\widetilde{\Delta}}^{R,A}_\sm{eq} =
    \nonumber \\
    &=& \tilde{\vare}^{R,A} \hat{\tau}_3 - D^{R,A}_\sm{eq} \hat{1} -
    \widehat{\widetilde{\Delta}}^{R,A}_\sm{eq} \,,
    \eea
depends only on the equilibrium self-energies,
($\widetilde{\vDelta}^{R,A}_\sm{eq} = \vDelta_\sm{eq}$ for a
triplet)
    \bea
    &&\tilde{\vare}^{R,A} = \vare - \Sigma^{R,A}_\sm{eq} - \sigma_B \,,
    \\
    &&D^{R,A}_\sm{eq} = n_{imp} t^{R,A}_+\,,
    \\
    &&\Sigma^{R,A}_\sm{eq} = n_{imp} t^{R,A}_- \,,
    \\
    &&\widetilde{\Delta}^{R,A}_\sm{eq} = \Delta_\sm{eq} + n_{imp} t^{R,A}_\Delta \,.
    \eea
We parameterize the anomalous propagator  as
    \be
    \delta\whg^a = \left( \begin{array}{cc}
    g^a + g'^a & f^a i\sigma_2 \\
    i\sigma_2 \ul{f}^a & g^a - g'^a
    \end{array} \right) \,,
    \ee
and find the equations for the diagonal components, $g^a$ and
$g'^a$, from (\ref{eq:lina})
\begin{widetext}
    \bea
    &&-g^a D^a + g'^a (\tilde{\vare}^R - \tilde{\vare}^A)
    +{1\over2} f^a (\ul{\widetilde{\Delta}}^R_\sm{eq} - \ul{\widetilde{\Delta}}^A_\sm{eq} )
    +{1\over2} \ul{f}^a (\widetilde{\Delta}^R_\sm{eq} - \widetilde{\Delta}^A_\sm{eq} ) + i\vv_f \grad_x \, g^a = 0 \,,
    \\
    &&-g'^a D^a + g^a (\tilde{\vare}^R - \tilde{\vare}^A)
    -{1\over2} f^a (\ul{\widetilde{\Delta}}^R_\sm{eq} + \ul{\widetilde{\Delta}}^A_\sm{eq} )
    +{1\over2} \ul{f}^a (\widetilde{\Delta}^R_\sm{eq} + \widetilde{\Delta}^A_\sm{eq} ) + i\vv_f \grad_x \, g'^a =
    -i\vv_f \grad_T \, \Phi_\sm{eq} \; (g_\sm{eq}^R - g_\sm{eq}^A) \,,
    \eea
where we defined $D^a = D^R_\sm{eq} - D^A_\sm{eq}$. Expressions
for the off-diagonal terms are obtained from the normalization
condition (\ref{eq:norma}),
    \bea
    f^a = {1\over g^R_\sm{eq} - g_\sm{eq}^A} (-g^a (f_\sm{eq}^R+f_\sm{eq}^A) +g'^a (f_\sm{eq}^R-f_\sm{eq}^A))
    \label{eq:fa} \\
    \ul{f}^a = {1\over g^R_\sm{eq} - g_\sm{eq}^A} (g^a (\ul{f}_\sm{eq}^R+\ul{f}_\sm{eq}^A) +g'^a (\ul{f}_\sm{eq}^R-\ul{f}_\sm{eq}^A))
    \label{eq:ffa}
    \eea
Combining these equations, and using the BPT approximation (i.e.
assuming that $g^a, g'^a = $const and spatially averaging the
terms containing $f$, $\ul{f}$ and $\Delta$), we obtain the final
expression for the unit-diagonal part of the anomalous propagator,
    \be
    g^a = \frac{G_2}{G_1 G_4-G_2 G_3} (g_\sm{eq}^R - g_\sm{eq}^A) (i\vv_f \grad \Phi_\sm{eq}) \,,
    \ee
with the following definitions of coefficients,
    \bea
    G_1 &=& -D^a + \frac{1}{2(g_\sm{eq}^R - g_\sm{eq}^A)} \, \overline{[
    -(f_\sm{eq}^R + f_\sm{eq}^A) (\ul{\widetilde{\Delta}}^R_\sm{eq} - \ul{\widetilde{\Delta}}^A_\sm{eq})
    +(\ul{f}_\sm{eq}^R + \ul{f}_\sm{eq}^A) (\widetilde{\Delta}^R_\sm{eq} - \widetilde{\Delta}^A_\sm{eq})]} \\
    G_2 &=& \tilde{\vare}^R - \tilde{\vare}^A + \frac{1}{2(g_\sm{eq}^R - g_\sm{eq}^A)} \, \overline{[
    (f_\sm{eq}^R - f_\sm{eq}^A) (\ul{\widetilde{\Delta}}^R_\sm{eq} - \ul{\widetilde{\Delta}}^A_\sm{eq})
    +(\ul{f}_\sm{eq}^R - \ul{f}_\sm{eq}^A) (\widetilde{\Delta}^R_\sm{eq} - \widetilde{\Delta}^A_\sm{eq})]} \\
    G_3 &=& \tilde{\vare}^R - \tilde{\vare}^A + \frac{1}{2(g_\sm{eq}^R - g_\sm{eq}^A)} \, \overline{[
    (f_\sm{eq}^R + f_\sm{eq}^A) (\ul{\widetilde{\Delta}}^R_\sm{eq} + \ul{\widetilde{\Delta}}^A_\sm{eq})
    +(\ul{f}_\sm{eq}^R + \ul{f}_\sm{eq}^A) (\widetilde{\Delta}^R_\sm{eq} + \widetilde{\Delta}^A_\sm{eq})]} \\
    G_4 &=& -D^a + \frac{1}{2(g_\sm{eq}^R - g_\sm{eq}^A)} \, \overline{[
    -(f_\sm{eq}^R - f_\sm{eq}^A) (\ul{\widetilde{\Delta}}^R_\sm{eq} + \ul{\widetilde{\Delta}}^A_\sm{eq})
    +(\ul{f}_\sm{eq}^R - \ul{f}_\sm{eq}^A) (\widetilde{\Delta}^R_\sm{eq} + \widetilde{\Delta}^A_\sm{eq})]} \,.
    \eea
In order to prove that our assumption of $\delta\whs^a=0$ is
justified, we note that the coefficients $G_i$ are even under
inversion of $\hvp$. Then the resulting $g^a$ and $g'^a$ are odd
due to additional factor $\vv_f \grad \Phi_\sm{eq}$, and their
averages over the Fermi surface vanishes. The same is true for the
off-diagonal functions $f^a$ and $\ul{f}^a$ in singlet case. This
assumption is not valid for triplet pairing: even though  $g^a$
and $g'^a$ are odd under inversion of $\hvp$, we see from
Eqs.(\ref{eq:fa}) and (\ref{eq:ffa}) that $f^a$ and $\ul{f}^a$ are
even (since $f_\sm{eq}$'s are odd), and additional terms due to
$\delta\whs^a$ appear, making the self-consistent solution of the
equations more difficult. Exceptions to this statement exist for
certain order parameters and for special orientation of the
temperature gradient. For example, when $\grad T$ is applied in a
direction along which ${\bm \Delta}$ does not change, we find
$\langle f^a\rangle = 0$. Two obvious examples are a)
$\vDelta(p_z)$ and $\grad T$ is in the $xy$-plane and b)
$\vDelta(p_x, p_y)$ and $\grad T \, || \, z$. )

The resulting expression for the heat conductivity is
    \be
    {\kappa_{ij} \over T} =
    \frac{N_f}{4\pi} \int {d\vare\over T} \left( \frac{\vare^2}{T^2 \cosh^2(\vare/2 T)} \right)
    \int d\vp_f \, n_f(\vp_f) (v_{f,i}(\hat{\vp}) v_{f,j}(\hat{\vp}))
    \frac{G_2}{G_1 G_4-G_2 G_3} (g_\sm{eq}^R - g_\sm{eq}^A)\,.
    \label{eq:HCON}
    \ee
\end{widetext}
We checked that this expression for a uniform superconductor
agrees with the heat conductivity of Graf \et\cite{MJGraf:1996}.
This completes the derivation of the heat current. We remind the
readers that in the main text we drop the equilibrium subscript
`eq' to make expressions less cluttered.



\end{document}